\documentclass[]{sn-jnl}
\usepackage[super,sort&compress,comma]{natbib}
\usepackage{multibib}
\newcites{methods}{Methods References}

\usepackage{mathtools} 
\usepackage{bbold}    
\usepackage{braket}    
\usepackage{gensymb}   

\usepackage{graphicx}%
\usepackage{multirow}%
\usepackage{amsmath,amssymb,amsfonts}%
\usepackage{amsthm}%
\usepackage{mathrsfs}%
\usepackage[title]{appendix}%
\usepackage{xcolor}%
\usepackage{textcomp}%
\usepackage{manyfoot}%
\usepackage{booktabs}%
\usepackage{algorithm}%
\usepackage{algorithmicx}%
\usepackage{algpseudocode}%
\usepackage{listings}%
\usepackage{setspace}
\usepackage{lineno}
\usepackage{amsmath}
\usepackage{times}
\usepackage{float}
\usepackage{bm}
\usepackage[normalem]{ulem}

\usepackage{xspace}
\usepackage{caption}
\DeclareCaptionLabelSeparator{bar}{\xspace\rule{0.4pt}{1em}\xspace}
\DeclareCaptionType{extfigure}[Extended Data Fig.][List of Extended Data Figures]
\makeatletter
\renewcommand\@biblabel[1]{#1.}
\makeatother

\newcommand{\ute}{UTe\textsubscript{2}\xspace}

\newcommand{\sro}{Sr\textsubscript{2}RuO\textsubscript{4}\xspace}

\newcommand{\Tc}{\ensuremath{T_{\rm c}}\xspace}
\newcommand{\Tco}{\ensuremath{T_{\rm c1}}\xspace}
\newcommand{\Tct}{\ensuremath{T_{\rm c2}}\xspace}
\newcommand{\Tcstar}{\ensuremath{T_{\rm c2}^*}\xspace}

\newcommand{\ctt}{\ensuremath{c_{33}}\xspace}
\newcommand{\coo}{\ensuremath{c_{11}}\xspace}
\newcommand{\cff}{\ensuremath{c_{55}}\xspace}
\newcommand{\att}{\ensuremath{\alpha_{33}}\xspace}
\newcommand{\Ps}{\ensuremath{P^{\star}}\xspace}
\newcommand{\Ts}{\ensuremath{T^{\star}}\xspace}
\newcommand{\BPT}{\ensuremath{B}-\ensuremath{P}-\ensuremath{T}\xspace}
\newcommand{\BbPT}{\ensuremath{B_b}-\ensuremath{P}-\ensuremath{T}\xspace}

\newenvironment{natureabstract}{%
	\par\vspace{0.25\baselineskip}%
	\noindent\begingroup
	\doublespacing            
	\bfseries                 
	\small                    
	{\large\bfseries Abstract}\quad    
}{%
	\par\endgroup\vspace{0.75\baselineskip}%
}

\unnumbered

\begin{document}
	
	\title{Thermodynamic demonstration of tetracriticality and multicomponent superconductivity in \ute}
	
	\author[1]{\fnm{Sahas} \sur{Kamat}}\nomail
	\author[2]{\fnm{Jared} \sur{Dans}}\nomail
	\author[2]{\fnm{Shanta} \sur{Saha}}\nomail
	\author[3]{\fnm{Artem D.} \sur{Kokovin}}\nomail
	\author[2,4]{\fnm{Johnpierre} \sur{Paglione}}\nomail
	\author[3,5]{\fnm{J\"org} \sur{Schmalian}}\nomail
	\author*[1,4]{\fnm{B. J.} \sur{Ramshaw}}\email{bradramshaw@cornell.edu}
	
	\affil*[1]{\orgdiv{Laboratory of Atomic and Solid State Physics}, \orgname{Cornell University}, \orgaddress{\city{Ithaca}, \postcode{14853}, \state{New York}, \country{United States}}}
	
	\affil[2]{\orgdiv{Maryland Quantum Materials Center, Department of Physics}, \orgname{University of Maryland}, \orgaddress{\city{College Park}, \state{Maryland}, \postcode{20742}, \country{United States}}}
	
	\affil[3]{\orgdiv{Institute for Theory of Condensed Matter}, \orgname{Karlsruhe Institute of Technology}, \orgaddress{\city{Karlsruhe}, \postcode{76131}, \country{Germany}}}
	
	\affil*[4]{\orgname{Canadian Institute for Advanced Research}, \orgaddress{\city{Toronto}, \state{Ontario}, \country{Canada}}}
	
	\affil[5]{\orgdiv{Institute for Quantum Materials and Technologies}, \orgname{Karlsruhe Institute of Technology}, \orgaddress{\city{Karlsruhe}, \postcode{76131}, \country{Germany}}}

	\maketitle

	\begin{natureabstract}
		
		The candidate topological superconductor \ute exhibits a complex phase diagram with multiple superconducting states, and the nature of their coexistence is not yet resolved. In particular, the apparent observation of two intersecting second-order phase boundaries at a triple point in the pressure-temperature phase diagram is thermodynamically forbidden. This suggests either a hidden phase transition or a fundamental misunderstanding of the superconductivity in this material. Here, we observe an additional phase boundary that is characterized by a upward jump in the sound velocity. This is direct thermodynamic evidence for a phase transition where superconducting order is lost upon cooling. Our results establish a tetracritical point beyond which the ambient and pressure-induced superconducting order parameters form a multi-component state. We use the measured phase diagram to demonstrate that strong competition between the two superconducting order parameters drives the loss of order when cooling, and leads to phase locking that suppresses superconducting fluctuations. These findings provide the definitive magnetic field-temperature-pressure phase diagram of \ute and establish a thermodynamic foundation for multi-component superconductivity. 
		
	\end{natureabstract}
	
	\section{Introduction}
	
	Multi-component superconductivity occurs when two (or more) order parameters condense: the most famous example---once thought to be the superconducting state of \sro\cite{kallinChiralPwaveOrder2012}---is the chiral $p_x+ip_y$ state in $^3$He \cite{vollhardtSuperfluidPhasesHelium2013}. Multi-component order parameters are a frequent ingredient for  chiral topological superconductivity \cite{sigristPhenomenologicalTheoryUnconventional1991a,satoTopologicalSuperconductorsReview2017}, but to date there are very few---if any---confirmed examples outside of $^3$He.
	
	\ute currently stands as the most viable candidate for spin-triplet, multi-component superconductivity. While it is generally agreed that the zero-field, ambient-pressure state (SC1) is single-component \cite{theussSinglecomponentSuperconductivityUTe22024,rosaSingleThermodynamicTransition2022} there is clear evidence for a distinct superconducting state (SC2) that emerges under hydrostatic pressure \cite{braithwaiteMultipleSuperconductingPhases2019,thomasEvidencePressureinducedAntiferromagnetic2020,wuMagneticSignaturesPressureInduced2025,vasinaConnectingHighFieldHighPressure2025}. Whether these phases coexist in some region of the pressure-temperature phase diagram and form a multi-component state is currently unknown.

	\begin{figure}[h]
		\centering
		\includegraphics[width=0.95\textwidth]{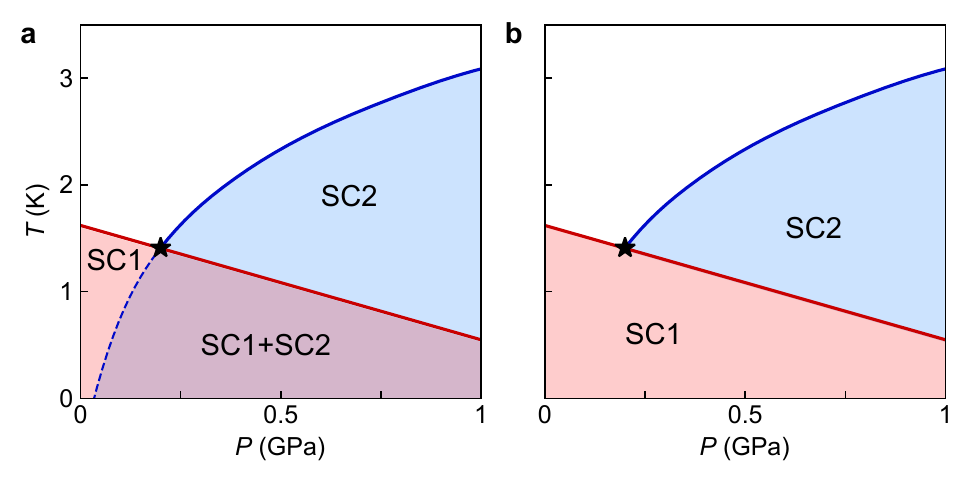}
		\caption{\textbf{Proposed temperature-pressure phase diagrams for \ute.} 
			The elusive multicritical point is indicated by a black star. The well-established SC1 and SC2 phase boundaries are shown as solid red and blue lines respectively. a) A proposed phase diagram where the SC2 phase boundary extends past the critical point, with a region of co-existence of the SC1 and SC2 phases (shaded purple). Such a scenario implies the existence of a phase boundary indicated with a blue dashed line---this boundary has not yet been observed. b) In a second proposal, the SC2 phase boundary terminates upon meeting the SC1 phase boundary. Given the weak heat capacity anomalies, this requires a third-order superconducting transition, where second derivatives of the free energy (such as heat capacity) remain continuous \cite{yipThermodynamicConsiderationsPhase1991,braithwaiteMultipleSuperconductingPhases2019}.}
		\label{fig:intro}
	\end{figure}
	
	
	The central challenge is the presence of an enigmatic multicritical point at $\left(\Ps,\Ts\right)$ in the pressure–temperature phase diagram, where the SC2 critical-temperature line terminates on the SC1 line (indicated by a star in \autoref{fig:intro}).
	Specific-heat \cite{braithwaiteMultipleSuperconductingPhases2019} and magnetization \cite{wuMagneticSignaturesPressureInduced2025} measurements clearly indicate that all transitions are second order and occur in the bulk.
	However, a “triple point” of second-order phase transitions, shown in Fig.~\ref{fig:intro}(b), is thermodynamically forbidden~\cite{yipThermodynamicConsiderationsPhase1991}, whereas the additional second-order line required for a tetracritical point, shown in Fig.~\ref{fig:intro}(a), has remained elusive.
	Attempts to resolve this puzzle have led to speculations ranging from a hidden phase boundary \cite{thomasEvidencePressureinducedAntiferromagnetic2020,vasinaConnectingHighFieldHighPressure2025,braithwaiteMultipleSuperconductingPhases2019,wuMagneticSignaturesPressureInduced2025} to a fine-tuned superconducting transition with a vanishing specific-heat jump \cite{vasinaConnectingHighFieldHighPressure2025,braithwaiteMultipleSuperconductingPhases2019}, suggesting the possibility of a third-order phase transition at \Ps~\cite{wuMagneticSignaturesPressureInduced2025}.

	
	Resolving these issues has implications that go beyond simply determining the topology of the \ute phase diagram: for example, the continuation of the SC2 phase boundary past the critical point, as sketched in \autoref{fig:intro}a, implies a region where the order parameters of SC1 and SC2 are simultaneously finite and with the potential for chiral topological superconductivity.

	In this work, we use pulse-echo ultrasound to map the field-pressure-temperature (\BPT) phase diagram of \ute in the vicinity of the critical pressure \Ps, where the two transitions meet. We find that the SC2 phase boundary extends past the SC1 phase boundary, leading to a region of two-component order SC1+SC2 in the phase diagram. We demonstrate that SC1 and SC2 interact strongly in this regime, with their competition leading to a loss of SC2 order upon cooling. We construct a phenomenological Ginzburg-Landau (GL) theory that accurately reproduces the observed phase diagram, and draw inferences from the constraints that our data place on the GL parameters and hence on the microscopic theory of superconductivity in \ute. 
	
	Finally, we explore the magnetic field dependence of this phase diagram and construct a coherent picture of the field-induced and pressure-induced superconducting states. We show that the multi-component phase is stabilized by a magnetic field and is accessible even at ambient pressure. Symmetry analysis of our \cff shear modulus data allows us to rule out the $A_u+B_{2u}$ and $B_{1u}+B_{3u}$ order parameter combinations, whether chiral or nonchiral, as candidates for the SC1+SC2 phase. Finally, the presence of a peak in the longitudinal sound attenuation at both \Tco\ and \Tct\ provides strong evidence that both SC1 and SC2 have sign-changing order parameters.
	
	\section{Experiment}
	Elastic moduli are thermodynamic coefficients: they are the second derivatives of a system's free energy with respect to strain. In this way, they are similar to the specific heat---the second derivative of the free energy with respect to temperature. Like specific heat, we expect non-analyticity in the elastic moduli at a phase transition. Unlike specific heat---where there is only one coefficient---we expect different behaviour for the different elastic moduli depending on how they couple to the order parameter \cite{ghoshThermodynamicEvidenceTwocomponent2021}. Generically, we expect jumps (discontinuities in elastic moduli) when the coupling is linear in strain, quadratic in order parameter, and kinks (discontinuities in the temperature derivatives of the elastic moduli) when the coupling is quadratic in strain, quadratic in order parameter. 
	
	Previous ultrasound measurements on \ute have demonstrated that the ambient and pressure-induced superconducting states couple strongly to the \coo and \ctt (compressional) and \cff (shear) moduli \cite{kamatVanishingPhaseStiffness2026,theussSinglecomponentSuperconductivityUTe22024}. We measure \coo,\ \ctt and \cff as a function of temperature at fixed pressures between $P=0$ and $P=1$ GPa, focusing on pressures near the critical point at $\Ps=0.20$ GPa (see \autoref{fig:intro}). Full experimental details are given in Kamat \textit{et al.} \cite{kamatVanishingPhaseStiffness2026} and in the Methods.

	\section{Results}
	
	\subsection{Elastic Moduli in Zero Magnetic Field}
	
			\begin{figure*}[h]
		\centering
		\includegraphics[width=\textwidth]{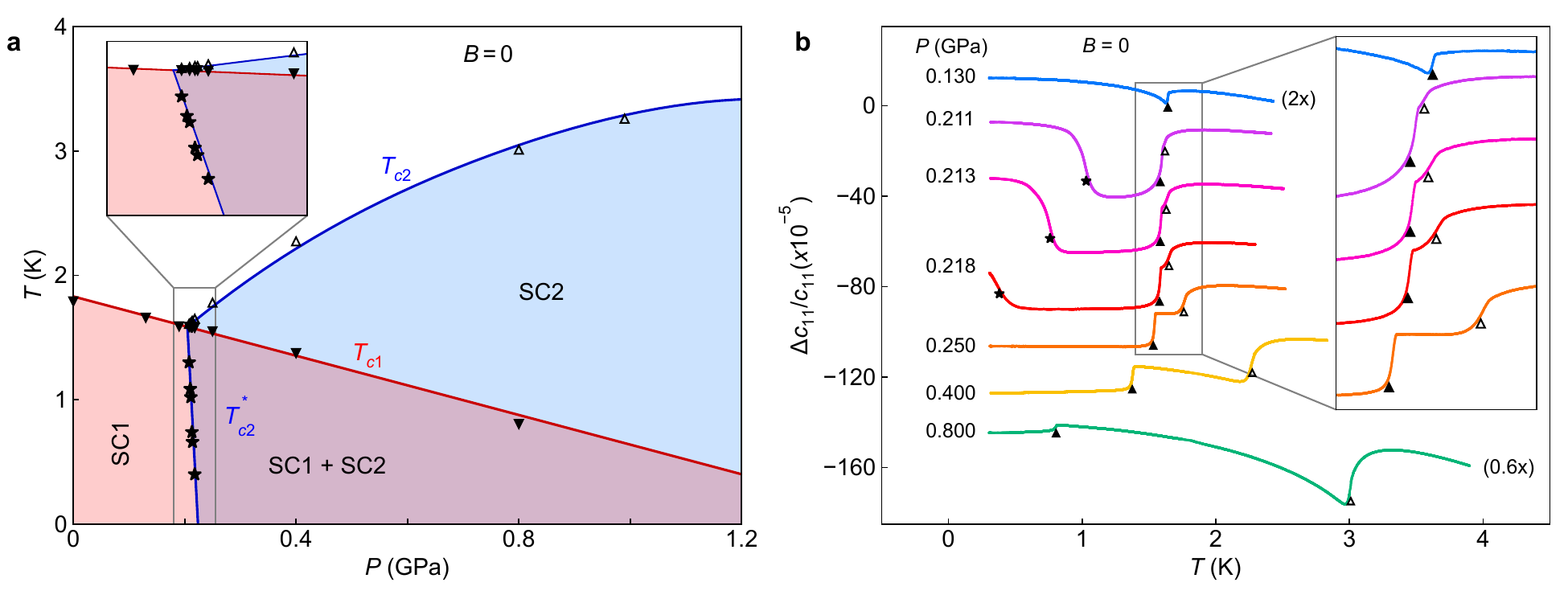}
		\caption{\textbf{The phase diagram of \ute near the tetracritical point.} a) The zero-field phase diagram of \ute constructed in this work. \Tco, \Tct and \Tcstar are represented by solid triangles, hollow triangles, and stars respectively. The solid red and blue lines are guides to the eye that represent the SC1 and SC2 phase boundaries, respectively. Regions in phase space where the SC1 and SC2 order parameters are non-zero are shaded red and blue, respectively. Inset: The phase diagram zoomed in around the tetracritical point, illustrating the back-bending of the \Tcstar phase boundary. b) Selected \coo elastic modulus data (full datasets are provided in the \hyperref[sec:extended]{Extended Data}). For $P<0.20$ GPa, we find a downward jump at \Tco when \ute enters the SC1 state. Two distinct jumps at at \Tco and \Tct are visible for $P>0.20$ GPa. Between $P=0.20$ and $0.22$ GPa, we find an upward jump at \Tcstar as the sample exits the SC2 superconducting state upon cooling. Inset: Zoomed in \coo data showing two distinct jumps at \Tco and \Tct for pressures near \Ps. }
		\label{fig:zerofield}
	\end{figure*}

	Measurements of the \coo compressional elastic modulus at different pressures and in zero magnetic field are shown in \autoref{fig:zerofield}b. At the lowest pressure---0.13 GPa---the transition to the SC1 state appears as a downward jump at \Tco. At pressures greater than \Ps, two jumps appear at two transitions whose separation in temperature increases with increasing pressure. We identify the upper and lower transitions as \Tct and \Tco, respectively, based on previous measurements of the phase diagram \cite{braithwaiteMultipleSuperconductingPhases2019,vasinaConnectingHighFieldHighPressure2025,kamatVanishingPhaseStiffness2026}. From $P=0.20$ to $0.22$ GPa ---pressures slightly greater than \Ps---we measure an additional feature in \ctt: an upward jump in \coo. The temperature at which this upward jump occurs does not correspond to either \Tco or \Tct, and we refer to it as \Tcstar.

	We use the thermodynamic signatures in the elastic moduli at \Tco and \Tct to construct the phase diagram shown in \autoref{fig:zerofield}a. These phase boundaries are consistent with previous work \cite{braithwaiteMultipleSuperconductingPhases2019,vasinaConnectingHighFieldHighPressure2025,kamatVanishingPhaseStiffness2026,thomasEvidencePressureinducedAntiferromagnetic2020,wuMagneticSignaturesPressureInduced2025}. For reasons we will justify in the remainder of this manuscript, we place the \Tcstar points on a phase boundary connected to \Tct at the critical pressure \Ps. This \Tcstar phase boundary has not been reported previously, and is the primary discovery of this work.
	
	\subsection{Loss of SC2 on Cooling}
	
	The phase diagram in \autoref{fig:zerofield}a suggests an unusual phenomenon: from $P = 0.20$ to $0.22$ GPa, when decreasing the temperature from above the upper transition, \ute first enters the SC2 phase at \Tct, then enters a mixed SC1+SC2 phase at \Tco, then loses the SC2 component of the order parameter cooling through \Tcstar. While the kinks and jumps in the elastic moduli identify the locations of these phase boundaries, we have not yet justified the assignment of \Tcstar to a phase boundary that connects to \Tct, nor have we justified the label SC1+SC2.
	
	Second-order phase transitions are defined by discontinuities in certain derivatives of the free energy \cite{callenThermodynamicsIntroductionThermostatistics1991}. The most familiar example is the specific heat, $C/T = -\partial^2 F/\partial T^2$. Typically, this jump must be upwards on cooling through \Tc as the system enters the ordered state. In a superconductor, opening a gap restricts the electronic degrees of freedom \cite{tinkhamIntroductionSuperconductivity2004}, causing the entropy decrease more rapidly below \Tc than in the normal state and producing the characteristic upward jump. For a phase boundary like \Tcstar, however, the heat capacity must jump downwards---behavior that is generally required at a back-bent phase transition line (see \hyperref[sec:methods]{Methods}). This has immediate implications for the elastic moduli.
	
	Like the specific heat, elastic moduli are second derivatives of the free energy: $c_{ijkl} = \partial^2 F / \partial \varepsilon_{ij}\partial \varepsilon_{kl}$. For compression strains (such as $\epsilon_{xx}$), the jumps in compression elastic moduli are related to the jump in the specific heat through an Ehrenfest relation. For $\coo \equiv c_{xxxx}$, the relationship is
	\begin{equation}
		\Delta \coo = -\frac{\Delta C}{\Tc} \left(\frac{\partial \Tc}{\partial\varepsilon_{xx}}\right)^2.
		\label{eq:ehrenfest}
	\end{equation}
	The coefficient $\left(\partial \Tc/\partial\epsilon_{xx}\right)$---the slope of \Tc as a function of strain---always enters as the square, which means that $\Delta \coo$ always has the opposite sign of $\Delta C$. Conventionally, this means that the jump in a compression modulus at \Tc must be downwards---a fact that is demonstrated at \Tco and \Tct in \autoref{fig:zerofield}c, and which has been documented for many other phase transitions \cite{rehwaldStudyStructuralPhase1973}.
	
	At $P=0.213$ GPa, \coo first jumps downward upon cooling through $\Tct=1.63$\ K, and again at $\Tco=1.58$\ K. Then, at $\Tcstar=1.0$\ K, \coo jumps upwards. By \autoref{eq:ehrenfest}, this requires that the specific heat jumps downwards, as required for a system that has lost order upon cooling. This is our primary evidence that \Tcstar marks the transition where the SC2 order parameter is lost upon cooling. We observe no hysteresis at \Tcstar when warming and cooling through the transition, demonstrating that this transition is second order --- see \hyperref[sec:methods]{Methods}, where we also show that such a transition is fully consistent with thermodynamic constraints.
	
	Having demonstrated that \Tcstar corresponds to a second order phase transition where order is lost upon cooling, we assign it to the second branch of the SC1 phase boundary. This boundary is nearly vertical in the $P-T$ phase diagram (\autoref{fig:zerofield}a). This assignment is consistent with the fact that the jump upwards at \Tcstar returns the elastic modulus to a value very close to what it would have been had it gone through \Tco alone, as can be determined from the 0.13 GPa data.
	
	
	
	\subsection{Elastic moduli in a magnetic field}
	\label{sec:infield}
	
	
			\begin{figure*}[h]
		\begin{center}
			\includegraphics[width=\textwidth]{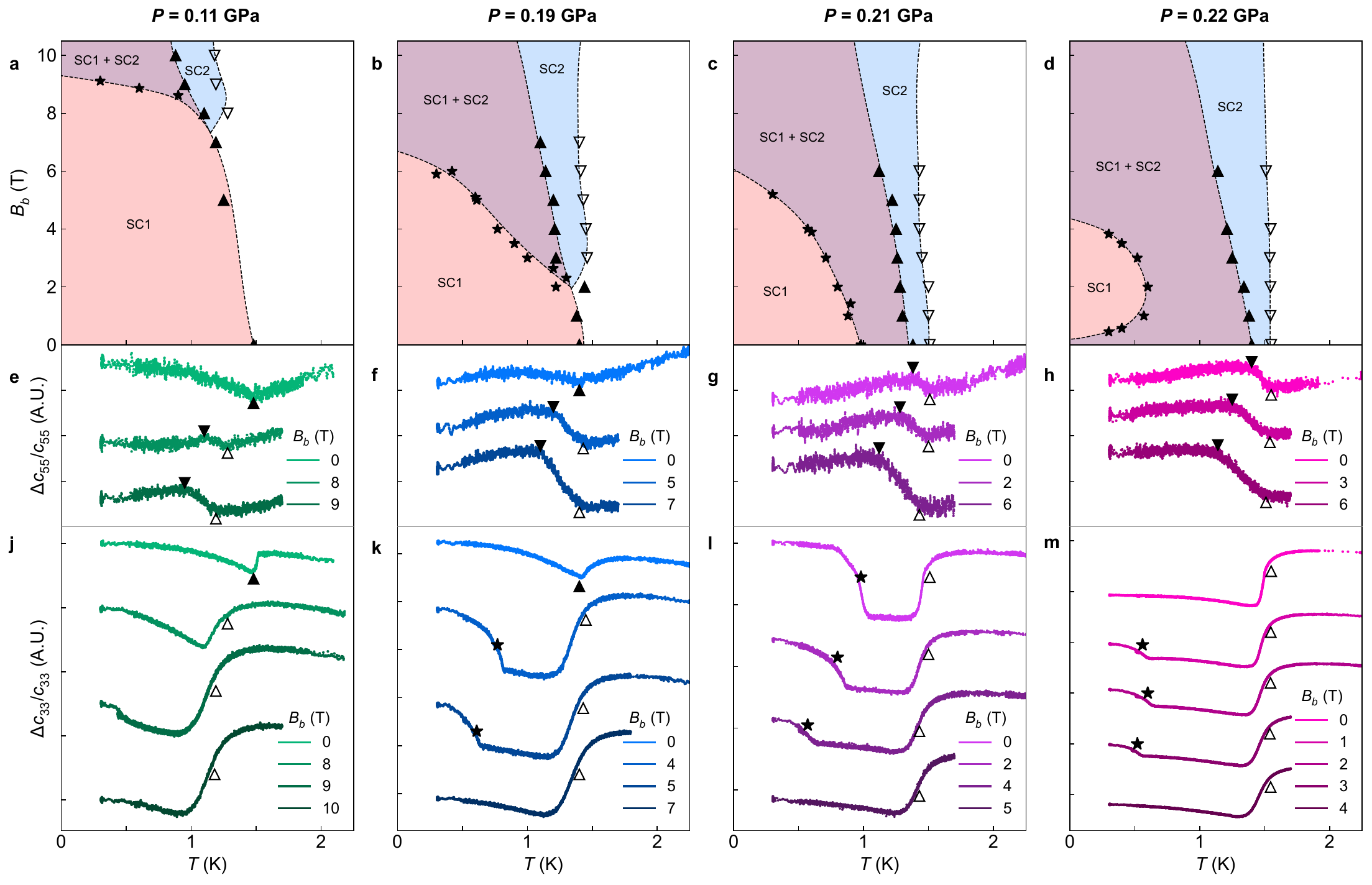}
		\end{center}
		\caption{\textbf{Field-temperature phase diagrams of \ute at fixed pressure}. Corresponding \cff and \ctt data are shown directly below. \Tco, \Tct and \Tcstar measured in this work are represented by solid triangles, hollow triangles, and stars respectively. a-d) Phase diagrams. Regions in phase space where the SC1, SC2, and SC1+SC2 order parameters are non-zero are shaded in red, blue, and purple, respectively. e-h) \cff elastic modulus data. \Tco and \Tct are seen as sharp kinks in the data. j-m) \ctt elastic modulus data. The downward jumps upon cooling correspond to \Tco or \Tct, while the upward jumps correspond to \Tcstar.
		}
		\label{fig:infield}
	\end{figure*}
	
	To explore the evolution of the \Tcstar phase boundary, we measure the \ctt and \cff elastic moduli as a function of magnetic field along the $b$-axis and track \Tcstar across the full \BbPT phase diagram.
	
	We construct field-temperature phase diagrams at 4 distinct pressures in the vicinity of \Ps, shown in \autoref{fig:infield}a-d. Below each phase diagram we show selected elastic moduli versus temperature curves that are used to construct these phase diagrams---full data sets are provided in the \hyperref[sec:extended]{Extended Data}. \autoref{fig:infield}e-h shows select \cff curves, and \autoref{fig:infield}j-m shows select \ctt curves. 
	
	
	We use kinks in \cff and jumps in \ctt to identify the phase boundaries: \Tco is indicated with solid triangles, and \Tct is indicated with hollow triangles. Note that \cff exhibits jumps only when the order paramters satisfy certain symmetry conditions \cite{theussSinglecomponentSuperconductivityUTe22024}; the absence of a jump in \cff here rules out the $A_u+B_{2u}$ and $B_{1u}+B_{3u}$ combinations, both chiral and nonchiral, for the SC1+SC2 state (see Methods). We find that the pressure-induced SC2 phase is indeed topologically connected to the field-induced phase for $B||b$ (not the orphan SC3 phase at higher fields away from the $b$ axis \cite{ranExtremeMagneticFieldboosted2019}), consistent with previous reports \cite{vasinaConnectingHighFieldHighPressure2025}.
	
	The central result contained in these data is the field and pressure evolution of \Tcstar. Similar to zero-pressure (but broadened slightly because of the magnetic field), we find upward jumps in \ctt upon cooling through \Tcstar. These phase boundaries represent the transition from the mixed SC1+SC2 phase to the pure SC1 phase---to our knowledge, these phase boundaries have never been reported. 
	
	
	The \Tcstar boundary in the field-temperature phase diagram evolves rapidly with pressure. At $P = 0.11$ GPa, the tetracritical point is located near 7 tesla and 1.25 kelvin. As pressure is increased, the field value at which this tetracritical point occurs decreases rapidly, reaching zero field at \Ps and providing the tetracritical point shown in \autoref{fig:zerofield}. Past \Ps, the pure SC1 phase is pinched off at zero magnetic field, confined to a small pocket at low temperatures and finite magnetic fields. 
	
	
	\subsection*{The full magnetic field-pressure-temperature phase diagram}
	\label{sec:tank}
	
	The phase diagrams we have constructed in $b$-axis magnetic field, pressure, and temperature, can be combined into a three-dimensional \BbPT diagram, shown in \autoref{fig:fishtank} (details of how this phase diagram is constructed are given in the caption). The two component state SC1+SC2 occupies a large region of phase space extending down to zero pressure for magnetic fields $18>B_b>12$ T. We show an ambient-pressure phase diagram as a function of $B_b$ in the \hyperref[sec:extended]{Extended Data}.

	\section{Analysis}
	
	\subsection{Competing superconducting orders}

	\begin{figure}[h]
		\begin{center}
			\includegraphics[width=0.8\textwidth]{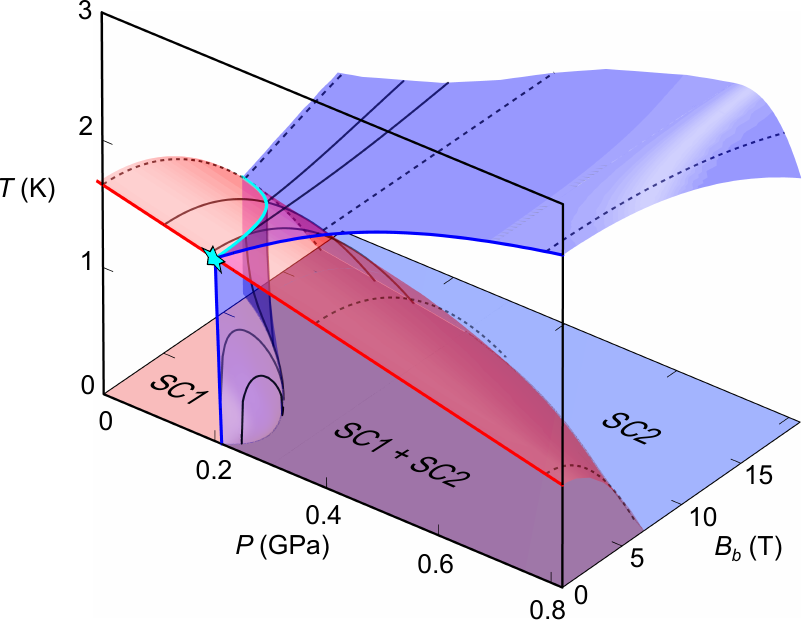}
		\end{center}
		\caption{\textbf{Field-Temperature-Pressure phase diagram of \ute.} The SC1 and SC2 phase boundaries are represented by red and blue sheets respectively. The $B=0$ plane is indicated by a black frame, on which solid red and blue lines represent the zero-field phase boundaries from this work (\autoref{fig:zerofield}a). Solid black lines represent the constant-pressure phase boundaries shown in \autoref{fig:infield}a-d. Dashed black lines are adapted from Vasina et. al. \cite{vasinaConnectingHighFieldHighPressure2025}. The zero temperature ground states of the system are shown on the $T=0$ plane. The tetracritical point in the $P-T$ plane---$\left(\Ps,\Ts\right)$, indicated with a light-blue star---becomes a line of tetracritical points in \BbPT space, indicated by the light-blue line. This line terminates at 12 tesla and 1 kelvin at ambient pressure. Between 12 and 18 tesla, the  SC1+SC2 state exists at ambient pressure (see \hyperref[sec:methods]{Methods} for an ambient-pressure phase diagram). }
		\label{fig:fishtank}
	\end{figure}
	
	The phase diagram shown in \autoref{fig:fishtank} is highly unusual in that the SC2 phase boundary bends back on itself when it hits the SC1 phase boundary. This leads to a narrow pressure range over which the SC2 order parameter is lost upon cooling: between 0.20 and 0.22 GPa, decreasing $T$ causes the system to first enter, and then exit the SC2 state. Can such an exotic phase diagram be explained within a Ginzburg Landau (GL) framework? While novel for a superconducting state, similar back-bending is observed at the intersection of antiferromagnetism (AFM) and superconductivity in the iron pnictides \cite{nandiAnomalousSuppressionOrthorhombic2010,fernandesCompetingOrderNature2010}. There, it was explained in terms of a competition between the superconducting and AFM order parameters (OPs) \cite{fernandesCompetingOrderNature2010}. In this section, we use the zero-field phase diagram to construct a phenomenological GL theory of two interacting superconducting OPs.

	We begin with two non-interacting OPs, $\psi_1$ and $\psi_2$, corresponding to the SC1 and SC2 states. Each independent $\psi_i$ has a free energy density $F_i=\frac{a_{i}}{2}\left|\psi_{i}\right|^{2}+\frac{u_{i}}{4}\left|\psi_{i}\right|^{4}$. Without any interaction between them, the total free energy is simply $F=F_1 + F_2$, with a phase diagram in the form of \autoref{fig:intro}a --- two phase boundaries crossing one another without any change in slope.
	
	However, the experimental SC2 phase boundary exhibits a significant change in slope when it hits the SC1 phase boundary. To account for this, we add the lowest-order, symmetry-allowed terms that couple $\psi_1$ and $\psi_2$:
	\begin{equation}
		F = F_1 + F_2 + \frac{\gamma_1}{2}|\psi_1|^2|\psi_2|^2 + \frac{\gamma_2}{4}(\psi_1^2\psi_2^{*2} + h.c.)
		\label{eq:GLtheory}
	\end{equation}
	Here, $\gamma_1$ couples the two OP magnitudes, while $\gamma_2$ locks the relative phase of the two complex OPs. The sign of $\gamma_2$ determines whether the SC1+SC2 breaks time reversal ($\gamma_2>0$) or not ($\gamma_2<0$). The competition of the two phases is governed in either case by the parameter $\gamma \equiv \gamma_1 - \left|\gamma_2\right|$.
	
	
	There are three key features of the experimental phase diagram that constrain the GL theory:
	(i) the phase lines meet at a tetracritical point;  (ii) the slope of \Tco versus pressure is almost unaffected by the intersection with \Tct; and (iii) the SC1 $\leftrightarrow$ SC1+SC2 phase boundary bends backwards. These three observations give rise to the following three constraints:
	\begin{equation}
		\begin{array}{c c}
			{\rm \underline{Observation}} &{\rm \underline {Constraint}} \\
			{\rm Tetracrit.\,\,point} & 0<\gamma<\sqrt{u_{1}u_{2}} \\
			T_{c1}{\rm \,unaffected} & \gamma\ll u_{2} \\
			{\rm Back\,\,bending} & \gamma > u_{1},
		\end{array}  
		\label{eq:cond}
	\end{equation}
	where the $u_i$ are the quartic coefficients of the uncoupled theory and we use a convention for the quadratic coefficients  $a_i$ discussed in \hyperref[sec:methods]{Methods}.

	\begin{figure*}[h]
		\centering
		\includegraphics[width=\textwidth]{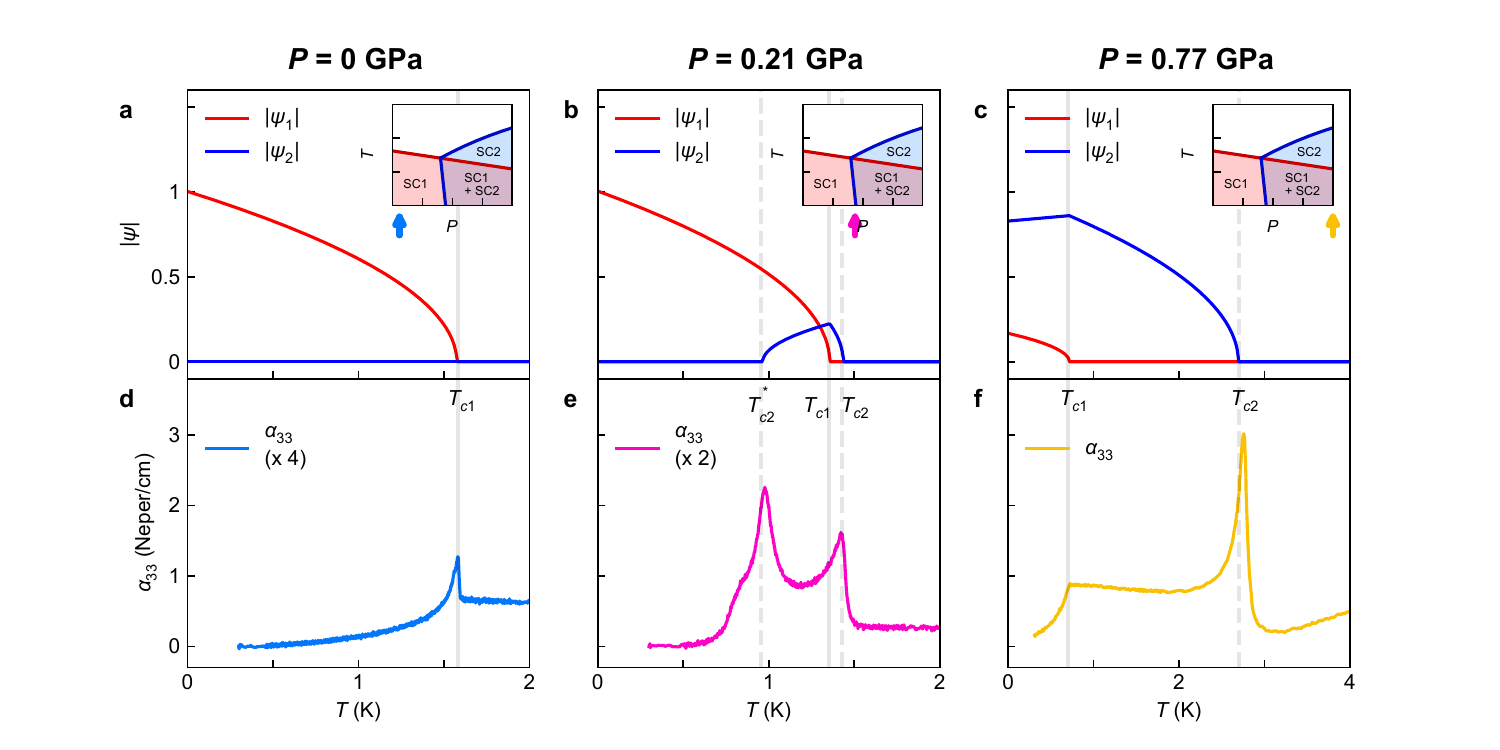}
		\caption{ \textbf{Order parameter competition in \ute.} a-c) The order parameters magnitudes for the SC1 ($|\psi_1|$) and SC2 ($|\psi_2|$) states versus temperature at different pressures. \Tco is indicated by a solid gray line, while \Tct and \Tcstar are indicated by dashed gray lines. Insets show where each pressure lies on the phase diagram of \autoref{fig:zerofield}a. d-f) Ultrasonic attenuation \att versus temperature at different pressures. We find an order-parameter-relaxation peak as the sample enters the SC1 state at 0 GPa, and as the sample enters and exits the SC2 state at 0.21 GPa. At 0.77 GPa, anomalously large phase fluctuations \cite{kamatVanishingPhaseStiffness2026} cause the ultrasonic attenuation inside the SC2 state to be larger than the normal state value. Below \Tco, \att is suppressed by the growth of the SC1 order parameter that provides phase stiffness and suppresses fluctuations.}
		\label{fig:OPs}
	\end{figure*}
	
	These constraints naturally explain the observed hierarchy of specific heat jumps at the various transitions. Specifically, they explain why the jump at the upper \Tct transition is small compared to the jump at \Tco---a fact that led to various alternative explanations of the phase diagram. First, the ratio of the specific heat jumps at the transitions from the normal metal to the single-component states is given by the inverse ratio of the quartic interactions,
	i.e. $(\left.\Delta C/T\right|_{T_{c1}})/(\left.\Delta C/T\right|_{T_{c2}}) =  u_{2}/u_{1}$. Eq.~\eqref{eq:cond} requires this ratio to be large---the jump at \Tco should be much larger than the jump a \Tct---which is fully consistent with the data of Refs.\cite{vasinaConnectingHighFieldHighPressure2025,braithwaiteMultipleSuperconductingPhases2019,thomasEvidencePressureinducedAntiferromagnetic2020}. Second, the jump at the transition from SC2 to SC2+SC1 is given by $\left.\Delta C/T\right|_{T_{c1}} = \left.\Delta C/T\right|_{T_{c2}}\times\frac{\left(u_{2}-\gamma\right)^{2}}{u_{1}u_{2}-\gamma^{2}}$.
	The above conditions require that the jump at \Tco to the SC1+SC2 phase is much larger than the jump at \Tct to the SC2 phase,
	which is again in full agreement with the data\cite{vasinaConnectingHighFieldHighPressure2025,braithwaiteMultipleSuperconductingPhases2019,thomasEvidencePressureinducedAntiferromagnetic2020}. Finally, a consequence of the back bending is that the jump at $T_{c2}^{*}$ is negative: $\left.\Delta C/T\right|_{T_{c2}^{*}}=-\left.\Delta C/T\right|_{T_{c1}}\times\frac{\left(u_{1}-\gamma\right)^{2}}{u_{1}u_{2}-\gamma^{2}}$, as discussed earlier. Full calculations are given in the \hyperref[sec:methods]{Methods}.
	
	With the GL parameters constrained by the phase diagram, the competition between SC1 and SC2 can be visualized by plotting $|\psi_1|$ and $|\psi_2|$ as a function of temperature alongside the \att sound attenuation coefficient. Sound attenuation in a metal at low temperatures is dominated by the conduction electrons---in the superconducting state, the sound attenuation generally decreases as the normal quasiparticles are gapped out (see Kamat \textit{et al.} \cite{kamatVanishingPhaseStiffness2026} for further discussion). \autoref{fig:OPs}a illustrates this at ambient pressure: the sound attenuation decreases to zero as $|\psi_1|$ grows with decreasing temperature. Near \Tco, we find an order parameter fluctuation peak in \att that is indicative of a sign-changing gap, as discussed in Kamat \textit{et al.} \cite{kamatVanishingPhaseStiffness2026} and seen in other heavy fermion superconductors \cite{bishopUltrasonicAttenuationUP$mathrmt_3$1984,batloggUltrasoundStudiesHeavy1985}.
	
	Competition between the two order parameters is clearly visible at $P = 0.21$ GPa---slightly greater than \Ps---as shown in \autoref{fig:OPs}b. The growth of $|\psi_2|$ below \Tct is sharply curtailed by the appearance of $|\psi_1|$ at \Tco. Below \Tco $\psi_1$ grows unimpeded, while $\psi_2$ is suppressed. This suppression is complete by \Tcstar. The ultrasonic attenuation \att, shown in \autoref{fig:OPs}e, shows attenuation peaks at both entrance to and exit out of the SC2 phase consistent with SC2 also having a sign-changing gap.
	
	At the highest pressure---$P = 0.77$ GPa---$|\psi_2|$ has a much broader region to establish itself before being truncated by the appearance of $|\psi_1|$ (\autoref{fig:OPs}c). The sound attenuation is most dramatic here: below the fluctuation peak at \Tct, the sound attenuation remains larger than its normal state value all the way down to \Tco, at which point it drops toward zero. We attribute this excess sound attenuation to the anomalously low phase stiffness of the SC2 phase \cite{kamatVanishingPhaseStiffness2026}---an attribution also consistent with the constraints in \autoref{eq:cond}, and with magnetic susceptibility measurements \cite{wuMagneticSignaturesPressureInduced2025}. Within our GL analysis, we show that the quartic coupling term $e(\psi_1^2\psi_2^{*2} + h.c.)$ locks the relative phase of the two OPs, allowing the SC2 state to gain phase stiffness from the SC1 state, suppressing the phase fluctuations that produce the anomalous sound attenuation for $\Tco<T<\Tct$ (see \hyperref[sec:methods]{Methods}). 
	
	\section{Discussion}
	
	Our discovery of the phase transition line at \Tcstar establishes that: 1) $\left(\Ps,\Ts\right) \approx \left(0.20 ~\rm{GPa}, 1.6~ \rm{K}\right)$ is a tetracritical point; 2) this extends to a tetracritical line in the \BPT phase diagram; 3) the SC2 phase is strongly influenced by the SC1 phase, but not the other way around; and 4) there is a region of the \BPT phase diagram with multi-component superconductivity. The obvious question is whether this multi-component state is time reversal symmetry breaking and possibly topological.
	
	Whether or not the SC1+SC2 state breaks time reversal depends on the sign of $\gamma_2$ in Eq.~\eqref{eq:GLtheory}. A negative value implies a trivial combination of the two OPs with the same phase, while a positive value implies that relative phase between the two OPs is $\pi/2$, leading to a chiral $\psi_1 + i\psi_2$ state that breaks time reversal symmetry (TRS). As we discuss in the \hyperref[sec:methods]{Methods}, if the spin structure  of the two triplet states in momentum space is similar, we find that $\gamma_2>0$ and that time-reversal symmetry breaking is favoured in SC1+SC2.
	
	An intriguing feature of the phase diagram is that the SC1+SC2 phase expands with increasing magnetic field (see \autoref{fig:fishtank}). Several mechanisms could account for this behavior. First, the field-induced expansion of the SC1+SC2 phase could arise from a time-reversal-symmetry-breaking multicomponent superconducting state. Such a state possesses an intrinsic TRS-antisymmetric orbital moment\cite{sigristPhenomenologicalTheoryUnconventional1991a} that couples linearly with a magnetic field, lowering the energy of the SC1+SC2 state. Second, it has recently been proposed by Helm \textit{et al.} \cite{helmFieldinducedCompensationMagnetic2024} that both the SC2 and high-field re-entrant (SC3) superconducting phases are stabilized by an internal exchange field via the Jaccarino--Peter effect, which could similarly favor the SC1+SC2 phase at finite field without the need for TRS breaking by the ordered state itself. Finally, given the proximity of SC2 to a metamagnetic phase transition \cite{rosuelFieldInducedTuningPairing2023,helmFieldinducedCompensationMagnetic2024,ranExtremeMagneticFieldboosted2019}, it is possible that the pairing interaction itself is strengthened by a magnetic field.
	
	The hierarchy of the quartic couplings, $u_{1}<\gamma\ll u_{2}$, has clear consequences for the electronic states that form the Cooper pairs. Such a pronounced separation of coupling strengths requires that the orbitals responsible for the large density of states---revealed by the residual specific heat and cyclotron mass---have a substantially larger weight in the pair wave function of SC1 than of SC2 (see \hyperref[sec:methods]{Methods}). This insight imposes a stringent constraint on any microscopic theory of the pairing mechanism in UTe$_{2}$.
	
	The accessibility of the multicomponent state at ambient pressure using a magnetic field motivates further study with probes directly sensitive to the excitations of a topological superconductor.
	
	\section{Funding Statement}
	Research at Cornell was supported by the Department of Energy, Office of Basic Energy Sciences Award No. DE-SC-0026003 (ultrasound measurements and data analysis). Research at the University of Maryland was supported by the Gordon and Betty Moore Foundation’s EPiQS Initiative Grant No. GBMF9071 (materials synthesis), the Department of Energy, Office of Basic Energy Sciences Award No. DE-SC-0019154 (sample characterization), the NIST Center for Neutron Research, and the Maryland Quantum Materials Center. Theory work was supported by the German Research Foundation TRR 288-422213477 ELASTO-Q-MAT,
	A07 (A.D.K. and J.S.) and grant  SFI-MPS-
	NFS-00006741-05 from the Simons Foundation (J.S.).
	
	\section{Author Contributions}
	B. J. R. and S.K. conceived the experiment. S.S. and J.P. grew and characterized samples of \ute. S.K. and J.D. prepared samples for ultrasound measurements and collected data. S.K.
	analyzed data and produced figures. J.S. and A.D.K. performed theoretical calculations. B.J.R. and S.K. wrote the manuscript, with input from all coauthors. B.J.R. supervised the project.
	
	\section{Competing Interests}
	The authors declare no competing interests.


	\section{Methods}
	\label{sec:methods}
	
	\subsection{Pressure determination}
	
	Samples were pressurized at room temperature in a piston anvil cell using a hydraulic press, with the resistance of a manganin coil mounted inside the pressure cell serving as a manometer. At cryogenic temperatures, pressure was determined by benchmarking measured \Tco and \Tct against existing literature \cite{braithwaiteMultipleSuperconductingPhases2019}\cite{vasinaConnectingHighFieldHighPressure2025}\cite{wuMagneticSignaturesPressureInduced2025}. We show the inferred pressures at low temperature ($P_{4K}$) versus the pressures measured by the manganin coil at room temperature ($P_{300K}$) in Extended Data Figure 1a. We obtain a good fit to a second order polynomial.
	
	To obtain a high degree of control over the relative pressure of our sample close to the tetracritical point, we tune pressure by carefully controlling the pressure cell volume. Measurements of the pressure as measured by the manganin coil versus the change in length of the pressure cell are shown in Extended Data Figure 1b. We find that the data fit well to a fourth order polynomial, with the observed curve reminiscent of a Boyle's law $P\propto1/V$ dependence. 
	
	Using the cryogenic pressure dependence we measure in Extended Data Figure 1a, we construct a function that relates the change in length of the cell ($\Delta L$) to the low-temperature pressure ($P_{4K}$). While the uncertainty in the fit in Extended Data Figure 1a means our absolute pressure determination is only accurate upto 0.05 GPa, we can accurately determine relative pressures around the tricritical point. Knowing the slope of the pressure versus length curve ($dP/dL$) allows us to tune the (relative) pressure of the cell by controlling its volume. 
	
	As a consistency check, we also measure \ctt from 4 to 100K at different pressures (Extended Data Figure 1d). The minimum in the data seen at temperatures near 15 K can be attributed to Kondo softening of the lattice. The magnitude of this dip is sensitive to changes in pressure, and the value of \ctt at the minimum decreases with increasing pressure. We observe this to be true at even the smallest pressure increases we measure in this work, with larger pressure increases leading to larger changes in the Kondo softening --- thus providing a qualitative consistency check to our pressure measurement procedure. 
	
	To observe discontinuities in thermodynamic quantities such as the elastic modulus or specific heat at \Tcstar, the pressure inhomogeniety in a measurement has to be smaller than the pressure-width of the phase boundary itself. Since the steep \Tcstar phase boundary is 0.01 GPa wide, any measurement of \Tcstar needs a smaller pressure inhomogenity. Our technique has an advantage in this regard because we measure a much smaller sample volume than other techniques such as specific heat (see Extended Data Figure 1e).
	
	\subsection{Phase Stiffness of the SC1 + SC2 state}
	
	In the Analysis section of the main text, we attribute the decrease in attenuation in the SC1+SC2 state to an increase in the phase stiffness as the SC1 order parameter grows below \Tco. Here, we sketch out a simple Ginzburg-Landau theory to show how the SC2 state acquires the phase stiffness of the SC1 state.
	
	In the SC2 state alone, the system is described by a free energy density ($F$) of the form
	\begin{equation}
		F = F_n + \int\!\! \frac{d\vec{r}}{V}~\left[\frac{a_2}{2}\left(T\right) \left| \psi_2 \right|^2 + \frac{u_2}{4} \left| \psi_2 \right|^4 + K_2 \left| \nabla\psi_2 \right|^2 \right]
		\label{eq:feng}
	\end{equation}
	Where $F_n$ is the free energy density of the normal state, and $a_2$, $u_2$ and $K_2$ are Ginzburg-Landau parameters corresponding to the SC2 state. The last (gradient) term in this equation penalizes spatial variation in the order parameter $\psi_2 = |\psi_2|e^{i\theta_2(x)}$ via an increase in the free energy density. For a sinusoidal spatial variation in the phase of the order parameter of the form $\theta_2(x) = \theta_0 e^{ikx}$, the increase in free energy density is $\Delta F = K_2 |\psi_2|^2 \left(k^2 \theta_0^2\right)$. The prefactor $\kappa = K_2 |\psi_2|^2$ is known as the phase stiffness. This quantity is anomalously low for the SC2 state, leading to large spatial fluctuations in $\theta_2(x)$ that cause a larger than normal state attenuation below \Tct \cite{kamatVanishingPhaseStiffness2026}. 
	
	When the sample enters the combined SC1+SC2 state, the total free energy density is given by:
	\begin{equation}
		\begin{aligned}
			F_{\text{tot}} =& F_n + \int\!\! \frac{d\vec{r}}{V}~\left[\frac{a_2}{2}\left| \psi_2 \right|^2 + \frac{u_2}{4} \left| \psi_2 \right|^4 + K_2 \left| \nabla\psi_2 \right|^2\right] \\&+ \int\!\! \frac{d\vec{r}}{V}~\left[\frac{a_1}{2} \left| \psi_1 \right|^2 + \frac{u_1}{4} \left| \psi_1 \right|^4 + K_1 \left| \nabla\psi_2 \right|^2\right] + \int\!\! \frac{d\vec{r}}{V}~\left[\frac{\gamma_2}{4}(\psi_1^2\psi_2^{*2} + \psi_1^{*2}\psi_2^2) + \frac{\gamma_1}{2}(|\psi_1|^2\psi_2|^2)\right] 
		\end{aligned}
		\label{eq:feng2}
	\end{equation}
	The term $\gamma_2(\psi_1^2\psi_2^{*2} + \psi_1^{*2}\psi_2^2)$ locks the relative phases of the two order parameters, such that if $\theta_2(x) = \theta_0 e^{ikx}$, $\theta_1$ is forced to be 
	\begin{equation}
		\theta_1(x) = \theta_0 e^{ikx} +
		\begin{cases}
			\pi/2 & e > 0 \\
			0 & e < 0 
		\end{cases}
	\end{equation}
	SC2 phase fluctuations thus lead to an increased energy cost by inducing fluctuations in the SC1 state.  The total increase in free energy density is given by
	\begin{equation}
		\Delta f = \left(K_1 |\psi_1|^2 + K_2 |\psi_2|^2 \right)\times\left(k^2 \theta_0^2\right)
	\end{equation}
	The total phase stiffness of the SC1+SC2 state becomes $\kappa=K_1 |\psi_1|^2 + K_2 |\psi_2|^2$. The SC1 state is known to have a larger phase stiffness, and $K_1 \gg K_2$. As $|\psi_1|$ grows below \Tco, $\kappa$ rapidly increases, suppressing phase fluctuations of the SC2 state and thus leading to a decreased attenuation. 
	
	\subsection{Ginzburg-Landau theory and constraints from the phase diagram}
	The Ginzburg-Landau expansion for a two-component superconductor with
	non-symmetry-related components, as it occurs for two distinct one-dimensional
	irreducible representations (irrep) is given as
	\begin{eqnarray}
		F\left(\psi_{1},\psi_{2}\right) & = & \frac{a_{1}}{2}\left|\psi_{1}\right|^{2}+\frac{u_{1}}{4}\left|\psi_{1}\right|^{4}+\frac{a_{2}}{2}\left|\psi_{2}\right|^{2}+\frac{u_{2}}{4}\left|\psi_{2}\right|^{4}\nonumber \\
		& + & \frac{\gamma_{1}}{2}\left|\psi_{1}\right|^{2}\left|\psi_{2}\right|^{2}+\frac{\gamma_{2}}{4}\left(\psi_{1}^{*}\psi_{1}^{*}\psi_{2}\psi_{2}+h.c.\right).\label{eq:GLtot}
	\end{eqnarray}
	where 
	\begin{equation}
		a_{j}=a_{j,0}\left(T-T_{c,j}^{\left(0\right)}\right).
	\end{equation}
	The $T_{c,j}^{\left(0\right)}$ are the transitions where the individual
	order parameters emerge first. The last term in Eq.(\ref{eq:GLtot})
	determines the relative phase $\phi_{1}-\phi_{2}$ of the two order
	parameters $\psi_{j}=A_{j}e^{i\phi_{j}}$. If $\gamma_{2}>0$, the
	relative phase is $\pm\pi/2$ and time-reversal symmetry (TRS) is
	broken if both order parameters occur simultaneously. If $\gamma_{2}$
	is negative TRS is preserved. Let us fix the global phase to be zero.
	Then it follows that 
	\begin{eqnarray}
		F\left(A_{1},A_{2}\right) & = & \frac{a_{1}}{2}A_{1}^{2}+\frac{u_{1}}{4}A_{1}^{4}+\frac{a_{2}}{2}A_{2}^{2}+\frac{u_{2}}{4}A_2^{4}\nonumber \\
		& + & \frac{\gamma}{2}A_{1}^{2}A_{2}^{2},
	\end{eqnarray}
	where $\gamma=\gamma_{1}-\left|\gamma_{2}\right|$. Let us further
	use $a_{j}=a_{j,0}\left(T-T_{c,j}^{\left(0\right)}\right)$. Assuming
	for example that $A_{2}\neq0$, we insert the stationary solution of the minimization
	w.r.t. $A_{2}$ and obtain an effective free energy for $A_{1}$:
	\begin{equation}
		F\left(A_{1}\right)=-\frac{a_{2}^{2}}{4u_{2}}+\frac{a_{1}}{2}\left(1-\frac{\gamma a_{2}}{a_{1}u_{2}}\right)A_{1}^{2}+\frac{u_{1}}{4}\left(1-\frac{\gamma^{2}}{u_{1}u_{2}}\right)A_{1}^{4}.
	\end{equation}
	Assuming the opposite switches $1\leftrightarrow2$. In both cases the new quartic term becomes negative if $\gamma^{2}>u_{1}u_{2}$ , signalling a first order transition. This corresponds to a bicritical
	point. If $\gamma^{2}<u_{1}u_{2}$ the transition is of second order and
	both the transitions meet at a tetracritical point, as shown in Extended Data Figure 2.
	
	Consider now the transition temperature of phase $1$ in the presence
	of order by phase 2. It holds 
	\begin{equation}
		T_{c,1}=T_{c,1}^{\left(0\right)}+\gamma\frac{a_{2,0}\left(T_{c,2}^{\left(0\right)}-T_{c,1}^{\left(0\right)}\right)}{\gamma a_{2,0}-u_{2}a_{1,0}}
	\end{equation}
	while in the opposite case $1\leftrightarrow2$.
	\begin{equation}
		T_{c,2}=T_{c,2}^{\left(0\right)}+\gamma\frac{a_{1,0}\left(T_{c,1}^{\left(0\right)}-T_{c,2}^{\left(0\right)}\right)}{\gamma a_{1,0}-u_{1}a_{2,0}}
	\end{equation}
	For $\gamma<\sqrt{u_{1}u_{2}}$, $T_{c,1}=T_{c,2}$ is only possible
	if $T_{c,1}^{\left(0\right)}=T_{c,2}^{\left(0\right)}$, which implies
	$T_{c,j}=T_{c,j}^{\left(0\right)}$. This is where the two bare transitions
	meet. 
	
	Let us now consider the case where the bare transition temperatures
	are controlled by pressure $p$ such that 
	\begin{eqnarray}
		T_{c,1}^{\left(0\right)} & = & T^{*}-\alpha_{1}\left(p-P^{*}\right),\nonumber \\
		T_{c,2}^{\left(0\right)} & = & T^{*}+\alpha_{2}\left(p-P^{*}\right),
	\end{eqnarray}
	where the $\alpha_{j}$ are assumed positive. The tetracritical point
	is then at $\left(T^{*},P^{*}\right)$. Consider now the regime $T_{c,1}<T_{c,2}$.
	Then follows 
	\begin{eqnarray}
		T_{c,1} & = & T^{*}-\alpha_{1}\left(p-P^{*}\right)\left(1-\gamma\frac{a_{2,0}\left(1+\alpha_{2}/\alpha_{1}\right)}{\gamma a_{2,0}-u_{2}a_{1,0}}\right)
	\end{eqnarray}
	When the slope of $T_{c,1}$ does not change significantly
	in the presence of phase $2$, it must hold 
	\begin{equation}
		\frac{\left(1+\alpha_{2}/\alpha_{1}\right)}{\left|1-u_{2}a_{1,0}/\gamma a_{2,0}\right|}\ll1.
	\end{equation}
	Suppose the two slopes $\alpha_{1}$ and $\alpha_{2}$ are of the same
	order of magnitude. It must hold that
	$u_{2}a_{1,0}$ is much larger than $\gamma a_{2,0}$.
	
	Consider now the opposite regime, where $T_{c,2}<T_{c,1}$. Then follows
	\begin{eqnarray}
		T_{c,2} & = & T^{*}+\alpha_{2}\left(p-p^{*}\right)\left(1-\gamma\frac{a_{1,0}\left(1+\alpha_{1}/\alpha_{1}\right)}{\gamma a_{1,0}-u_{1}a_{2,0}}\right)
	\end{eqnarray}
	The back bending of the phase line implies that $\frac{dT_{c,2}}{dp}<0$
	in the presence of order of $\psi_1$. Hence, it must hold that 
	\begin{equation}
		\gamma\frac{a_{1,0}\left(1+\alpha_{1}/\alpha_{1}\right)}{\gamma a_{1,0}-u_{1}a_{2,0}}>1
	\end{equation}
	A necessary condition for this to happen is that 
	\begin{equation}
		\gamma>\frac{a_{2,0}}{a_{1,0}}u_{1}
	\end{equation}
	The slope is particularly large, if $\gamma$ is only slightly larger
	than $\frac{a_{2,0}}{a_{1,0}}u_{1}$ and if $\gamma a_{1,0}$ is not
	too small. 
	
	To summarize, we have the following conditions:
	\begin{eqnarray}
		0 & < & \gamma<\sqrt{u_{1}u_{2}}\nonumber \\
		\gamma & \ll & \frac{a_{1,0}}{a_{2,0}}u_{2}\nonumber \\
		\gamma & > & \frac{a_{2,0}}{a_{1,0}}u_{1}
	\end{eqnarray}
	which follow from the existence of a tetracritical point,  the weak change in slope of $T_{c,1}$, and the back bending of $T_{c,2}$, as discussed in the main paper.
	
	In our free energy expansion we have a choice to properly normalize
	the order parameters. Hence, we can independently rescale the two
	order parameters,
	\begin{equation}
		\left(\begin{array}{c}
			\psi_{1}\\
			\psi_{2}
		\end{array}\right)\rightarrow\left(\begin{array}{c}
			s_{1}\psi_{1}\\
			s_{2}\psi_{2}
		\end{array}\right).
	\end{equation}
	This would not be allowed for a two-dimensional irrep, where the components
	transform into each other under certain symmetry operations and only
	$s_{1}=s_{2}$ is allowed. We can use this freedom to transform the
	quadratic terms such that $\left(a_{1,0},a_{2,0}\right)\rightarrow\left(s_{1}^{2}a_{1,0},s_{2}^{2}a_{2,0}\right)$.
	If one then choses $s_{2}=\sqrt{\frac{a_{1,0}}{a_{2,0}}}s_{1}$ it follows $a_{1,0}=a_{2,0}$. Then we obtain the
	constraints given in the main text.
	
	\subsubsection{Heat capacity jumps}
	
	The extra heat capacity (relative to the normal state) in a phase
	with $\psi_{1}\neq0$ and $\psi_{2}=0$ is 
	\begin{equation}
		\delta C_{1}=\frac{a_{1,0}^{2}T}{2u_{1}},
	\end{equation}
	while the opposite case ($\psi_{1}=0$ but $\psi_{2}\neq0$) yields
	\begin{equation}
		\delta C_{2}=\frac{a_{2,0}^{2}T}{2u_{2}},
	\end{equation}
	On the other hand, the heat capacity in the state where both order
	parameters are finite is 
	\begin{eqnarray}
		\delta C_{12} & = & \frac{a_{2,0}^{2}u_{1}+a_{2.0}^{2}u_{2}-2a_{1,0}a_{2,0}\gamma}{2\left(u_{1}u_{2}-\gamma^{2}\right)}T.\nonumber \\
		& = & \frac{\left(a_{2,0}\sqrt{u_{1}}-a_{2.0}\sqrt{u_{2}}\right)^{2}+2a_{1,0}a_{2,0}\left(\sqrt{u_{1}u_{2}}-\gamma\right)}{2\left(u_{1}u_{2}-\gamma^{2}\right)}T.
	\end{eqnarray}
	The last line merely shows that $\delta C_{12}>0$.
	
	Hence, the jumps of the heat capacity at the upper, single transitions
	are 
	\begin{equation}
		\frac{\Delta C\left(T_{c,1}^{\left(0\right)}\right)}{T_{c,1}^{\left(0\right)}}=\frac{a_{1,0}^{2}}{2u_{1}}\,{\rm and\,}\frac{\Delta C\left(T_{c,2}^{\left(0\right)}\right)}{T_{c,2}^{\left(0\right)}}=\frac{a_{2,0}^{2}}{2u_{2}}
	\end{equation}
	The jump from a phase with ($\psi_{1}=0$ and $\psi_{2}\neq0$) to
	a phase with ($\psi_{1}\neq0$ and $\psi_{2}\neq0$) at $T_{c,1}$
	we get 
	\begin{equation}
		\frac{\Delta C\left(T_{c,1}\right)}{T_{c,1}}=\frac{\delta C_{12}\left(T_{c,1}\right)-\delta C_{2}\left(T_{c,1}\right)}{T_{c,1}}=\frac{\left(a_{1,0}u_{2}-a_{2,0}\gamma\right)^{2}}{2u_{2}\left(u_{1}u_{2}-\gamma^{2}\right)}.
	\end{equation}
	If we now analyze the ratio between the lower and upper transition,
	it follows
	\begin{equation}
		\frac{T_{c,2}^{\left(0\right)}\Delta C\left(T_{c,1}\right)}{T_{c,1}\Delta C\left(T_{c,2}^{\left(0\right)}\right)}=\frac{\left(\frac{a_{1,0}}{a_{2,0}}u_{2}-\gamma\right)^{2}}{u_{1}u_{2}-\gamma^{2}}.
	\end{equation}
	We can also analyze the jump from a phase with ($\psi_{1}\neq0$
	and $\psi_{2}\neq0$) to a phase with ($\psi_{1}\neq0$ but $\psi_{2}=0$)
	in the low-$T$ phase (because of the back-bending) at $T_{c,2}$:
	\begin{equation}
		\frac{\Delta C\left(T_{c,2}\right)}{T_{c,2}}=\frac{\delta C_{1}\left(T_{c,2}\right)-\delta C_{12}\left(T_{c,2}\right)}{T_{c,2}}=-\frac{\left(a_{2,0}u_{1}-a_{1,0}\gamma\right)^{2}}{2u_{1}\left(u_{1}u_{2}-\gamma^{2}\right)},
	\end{equation}
	i.e. the heat capacity should have a negative jump as one enters SC1
	through the back-bent phase transition. 
	
	\subsection{Microscopic derivation of Ginzburg-Landau coefficients}
	
	Let us next get a microscopic understanding of the coefficients of the Ginzburg-Landau theory. To this end we start from an effective Fermi liquid description in terms of well-defined quasi-particles, even if those excitations are the result of strong electronic interactions. Hence, we start from  a Boboliubov-de Gennes Hamiltonian 
	\begin{equation}
		\hat{H}_{\text{BdG}}=\sum_{\boldsymbol{k}}\left(\boldsymbol{c}_{\boldsymbol{k}}^{\dagger}\xi(\boldsymbol{k})\boldsymbol{c}_{\boldsymbol{k}}+\frac{1}{2}\boldsymbol{c}_{\boldsymbol{k}}^{\dagger}\Delta(\boldsymbol{k})\boldsymbol{c}_{-\boldsymbol{k}}^{\dagger T}+\frac{1}{2}\boldsymbol{c}_{\boldsymbol{-k}}^{T}\Delta^{\dagger}(\boldsymbol{k})\boldsymbol{c}_{\boldsymbol{k}}\right).
	\end{equation}
	As we want to achieve a description in terms of a multi-orbital problem with spin-orbit interactions, $c_{\boldsymbol{k}}$ is a vector of annihilation operator of the electrons
	in bands with different pseudo-spins. If we consider a system with inversion-
	and time-reversal symmetry in the normal state, each $\boldsymbol{k}$ state is doubly (Kramers)
	degenerate. Except for this Kramers degeneracy, the bands are for each $\boldsymbol{k}$ and in the generic case, non-degenerate in the vicinity of the Fermi surface. Hence,  there is a single and unique band that it belongs to a given Fermi momentum. This merely simplifies out notation and allows to use momentum as only relevant quantum number.
	Let us focus on the case where both order parameters are peudo-spin triplets, i.e. have a structure 
	\begin{equation}
		\Delta(\boldsymbol{k})=i\sigma_{y}\left(\psi_{1}\boldsymbol{\Delta}_{1}(\boldsymbol{k})+\psi_{2}\boldsymbol{\Delta}_{2}(\boldsymbol{k})\right)\cdot\boldsymbol{\sigma}.
	\end{equation}
	$\sigma$ are Pauli matrices, acting in the pseudo-spin space. The $\boldsymbol{\Delta}_{1,2}(\boldsymbol{k})$ are form factors of the pairing states that determine the spin and momentum structure of the Cooper pairs. For example, for a state that transforms under the $B_{1u}$ irrep
	holds 
	\begin{equation}
		\boldsymbol{\Delta}\left(\boldsymbol{k}\right)=\phi_{x}\boldsymbol{e}_{x}\sin k_{y} b +\phi_{y}\boldsymbol{e}_{y}\sin k_{x} a+\phi_{z}\boldsymbol{e}_{z}\sin\frac{k_{x} a}{2}\sin\frac{k_{y} b }{2}\sin\frac{k_{z} c}{2},
		\label{eq:B1u}
	\end{equation}
	where $a$, $b$ and $c$ are the unit cell lattice constants in the corresponding directions. Other irreps can easily be described accordingly.  These form factors of the pairing states require an appropriate normalization which we discuss below. Integrating out the fermions and expanding the free energy of the system in terms of the two order parameters $\psi_1$ and $\psi_2$ then yields some microscopic expressions for the parameters of the Ginzburg-Landau expansion.
	
	To express the coefficients of the energy expansions, it is useful to use the following notation:
	\begin{eqnarray}
		\braket{\boldsymbol{\Delta}_{i}|\boldsymbol{\Delta}_{j}}&=&\int\frac{dS}{|v(\boldsymbol{k})|}\boldsymbol{\Delta}_{i}^{*}(\boldsymbol{k})\cdot\boldsymbol{\Delta}_{j}(\boldsymbol{k}),\\ 
		\braket{\boldsymbol{\Delta}_{i}\otimes\boldsymbol{\Delta}_{k}|\boldsymbol{\Delta}_{j}\otimes\boldsymbol{\Delta}_{m}}&=&\int\frac{dS}{|v(\boldsymbol{k})|}\left(\boldsymbol{\Delta}_{i}^{*}(\boldsymbol{k})\cdot\boldsymbol{\Delta}_{j}(\boldsymbol{k})\right)\left(\boldsymbol{\Delta}_{k}^{*}(\boldsymbol{k})\cdot\boldsymbol{\Delta}_{l}(\boldsymbol{k})\right),
	\end{eqnarray}
	where integration is performed over the Fermi surface and states are weighted by the inverse Fermi velocity.  We then obtain for the corresponding coefficients in the Ginzburg-Landau expansion:
	\begin{gather}
		\left|\psi_{1}\right|^{2}:a_{1,0}=\frac{\braket{\boldsymbol{\Delta}_{1}|\boldsymbol{\Delta}_{1}}}{T_{c,1}^{(0)}},\\
		\left|\psi_{2}\right|^{2}:a_{2,0}=\frac{\braket{\boldsymbol{\Delta}_{2}|\boldsymbol{\Delta}_{2}}}{T_{c,2}^{(0)}},\\
		\left|\psi_{1}\right|^{4}:u_{1}=\frac{7\zeta(3)}{2\pi^{2}\left(T\right)^{2}}\left(2\braket{\boldsymbol{\Delta}_{1}\otimes\boldsymbol{\Delta}_{1}|\boldsymbol{\Delta}_{1}\otimes\boldsymbol{\Delta}_{1}}-\braket{\boldsymbol{\Delta}_{1}\otimes\boldsymbol{\Delta}_{1}^{*}|\boldsymbol{\Delta}_{1}^{*}\otimes\boldsymbol{\Delta}_{1}}\right),\\
		\left|\psi_{2}\right|^{4}:u_{2}=\frac{7\zeta(3)}{2\pi^{2}\left(T\right)^{2}}\left(2\braket{\boldsymbol{\Delta}_{2}\otimes\boldsymbol{\Delta}_{2}|\boldsymbol{\Delta}_{2}\otimes\boldsymbol{\Delta}_{2}}-\braket{\boldsymbol{\Delta}_{2}\otimes\boldsymbol{\Delta}_{2}^{*}|\boldsymbol{\Delta}_{2}^{*}\otimes\boldsymbol{\Delta}_{2}}\right),\\
		\left|\psi_{1}\right|^{2}\left|\psi_{2}\right|^{2}:\gamma_{1}=\frac{7\zeta(3)}{4\pi^{2}\left(T\right)^{2}}\left(4\braket{\boldsymbol{\Delta}_{1}\otimes\boldsymbol{\Delta}_{2}|\boldsymbol{\Delta}_{1}\otimes\boldsymbol{\Delta}_{2}}+4\braket{\boldsymbol{\Delta}_{2}\otimes\boldsymbol{\Delta}_{2}^{*}|\boldsymbol{\Delta}_{1}\otimes\boldsymbol{\Delta}_{1}^{*}}-4\braket{\boldsymbol{\Delta}_{1}^{*}\otimes\boldsymbol{\Delta}_{1}|\boldsymbol{\Delta}_{2}\otimes\boldsymbol{\Delta}_{2}^{*}}\right),\\
		\psi_{1}^{2}\psi_{2}^{*2}:\gamma_{2}=\frac{7\zeta(3)}{4\pi^{2}\left(T\right)^{2}}(2\braket{\boldsymbol{\Delta}_{2}\otimes\boldsymbol{\Delta}_{2}|\boldsymbol{\Delta}_{1}\otimes\boldsymbol{\Delta}_{1}}-\braket{\boldsymbol{\Delta}_{1}^{*}\otimes\boldsymbol{\Delta}_{2}|\boldsymbol{\Delta}_{1}\otimes\boldsymbol{\Delta}_{2}^{*}}),\\
		\gamma=\gamma_{1}-|\gamma_{2}|.
	\end{gather}
	
	These results allow for several rather general insights:
	\begin{itemize}
		\item The above freedom to rescale the order parameters is equivalent to choosing an appropriate normalization of $\boldsymbol{\Delta}_{1,2}(\boldsymbol{k})$. For example, if we adopt the natural normalization
		\begin{equation}
			\braket{\boldsymbol{\Delta}_{i}|\boldsymbol{\Delta}_{j}}=\int\frac{dS}{|v|}\boldsymbol{\Delta}_{i}^{\dagger}\cdot\boldsymbol{\Delta}_{j}=\delta_{ij}.\label{eq:normalization}
		\end{equation}
		we obtain $a_{i,0}=1/T_{c,i}^{(0)}$. Close to the tetra-critical point, where the two transitions merge, this is essentially equivalent to setting $a_{1,0}=a_{2,0}$.
		
		\item If we now ask how the hierarchy $u_1<\gamma\ll u_2$, which follows from the phase diagram, can be realized, it is not straightforward to identify two form factors $\boldsymbol{\Delta}_{1}(\boldsymbol{k})$ and $\boldsymbol{\Delta}_{2}(\boldsymbol{k})$ for the pairing states SC1 and SC2 that share the same normalization of $\boldsymbol{\Delta}^*_{i}(\boldsymbol{k})\cdot \boldsymbol{\Delta}_{i}(\boldsymbol{k})$, see Eq.~\eqref{eq:normalization}, yet whose square differs sufficiently to ensure $u_1\ll u_2$. For pairing states such as that of Eq.~\eqref{eq:B1u}, whose leading harmonics vary smoothly throughout the Brillouin zone, achieving this separation is difficult for Fermi surface sheets that contribute similarly to the density of states. However, in our normalization scheme, states are weighted by their inverse Fermi velocity, i.e. their contribution to the density of states. If the orbital weight of $\boldsymbol{\Delta}_{1,2}(\boldsymbol{k})$ differs significantly between bands with large and small Fermi velocities, this issue may be resolved and the observed phase diagram can be explained. We therefore conclude that it is rather likely that the superconducting states SC1 and SC2 differ substantially in their orbital composition on the heavy U-$5f$ bands, compared to bands with U-$6d$ or Te-$5p$ character. In particular, the larger value of $u_2$ implies a smaller contribution of U-$5f$ orbitals in SC2.
		
		\item The sign of the expansion coefficient $\gamma_2$ determines whether TRS is broken in the SC1+SC2 phase. If we consider so-called unitary superconducting states, the $\boldsymbol{\Delta}_{1,2}$ are real and we obtain
		\begin{equation}
			\gamma_{2}=\frac{7\zeta(3)}{4\pi^{2}\left(T\right)^{2}}\int\frac{dS}{|v(\boldsymbol{k})|}\left|\boldsymbol{\Delta}_{1}\right|^{2}\left|\boldsymbol{\Delta}_{2}\right|^{2}\left(\cos2\theta_{\boldsymbol{k}}\right),
		\end{equation}
		where $\theta_{\boldsymbol{k}}$ is the angle between $\boldsymbol{\Delta}_{1}(\boldsymbol{k})$ and $\boldsymbol{\Delta}_{2}(\boldsymbol{k})$. Thus, whenever the spin structure of the two triplet states is such that, upon averaging over all Fermi-surface sheets, one predominantly has $\theta_{\boldsymbol{k}}\sim 0$ or $\sim \pi$, i.e. the spin structures of the two states are aligned in momentum space, it follows that $\gamma_2>0$ and TRS is broken; otherwise TRS remains preserved. Given our above insights for the orbital composition of the pairing states, it seems natural that this spin alignment of the pairing states only has to take place on the FS sheets dominated by U-$5f$ states, since this FS sheet dominates the total sum over all states.
	\end{itemize}
	
	\subsection{General Thermodynamic Constraints}
	\label{sec:yip}
	
	In this section, we show that the phase diagram observed in this work (\autoref{fig:zerofield}a) is consistent with the constraints imposed by the measured specific heat. Following a similar analysis performed in UPt$_3$ by Yip et al. \cite{yipThermodynamicConsiderationsPhase1991}, and using their notation, the normal state, SC2, SC1 and SC1+SC2 phases are referred to as N, A, B and C respectively. The slopes $dP/dT$ of the $AN$, $BN$, $CA$ and $CB$ phase boundaries are defined to be $p_1$, $p_2$, $p_3$ and $p_4$ respectively (see Extended Data Figure 3). The difference in specific heat (denoted $\alpha$ in Yip \textit{et al.} \cite{yipThermodynamicConsiderationsPhase1991}) across any phase boundary $IJ$ is $\Delta \alpha_{IJ} \equiv \alpha_{I} - \alpha_{J}$. We also define ratios of the specific heat jumps:
	\[
	\begin{aligned}
		r_3 &= \left(\frac{\Delta \alpha_{CA}}{\Delta \alpha_{BN}}\right)^{1/2},
		\qquad\qquad
		r_1 &= \left(\frac{\Delta \alpha_{AN}}{\Delta \alpha_{BN}}\right)^{1/2}.
	\end{aligned}
	\]
	
	Since the line $CB$ is a second order phase boundary, Ehrenfest's relations require that \cite{yipThermodynamicConsiderationsPhase1991}
	\begin{equation}
		1-\frac{p_2}{p_3} = \left(\frac{p_2}{p_1} - 1\right)r_1\frac{r_1r_3 \pm (r_1^2+r_3^2 - 1)^{1/2}}{(1-r_1^2)r_3}.
		\label{eq:yip1}
	\end{equation}
	The positivity of the term in the square-root gives us our first constraint. Since $\Delta \alpha_{CB} = \Delta \alpha_{BN} (r_1^2 + r_3^2 -1)$, this requires $\Delta \alpha_{CB}>0$: i.e., when temperature is lowered across the $CB$ phase boundary, the heat capacity jumps down; or alternatively, the elastic constant jumps up. This is exactly what we observe. It is interesting to note that the authors of Yip \textit{et al.} \cite{yipThermodynamicConsiderationsPhase1991} use $\Delta \alpha_{CB}>0$ to conclude that $C$ must be a lower temperature phase than $B$, presumably owing to a lack of observed downward jumps in heat capacity at the time. We experimentally observe an upward jump in the elastic constant in this work, requiring a downward jump in the specific heat capacity to be consistent with Ehrenfest relations, therefore requiring $B$\ (i.e. SC1) to be the lower temperature phase. We show later in this section that this still allows all constraints to be satisfied.
	
	Defining
	\begin{equation}
		r_4 \equiv \left(\frac{\Delta \alpha_{CB}}{\Delta \alpha_{BN}}\right)^{1/2} = (r_1^2 + r_3^2 -1)^{1/2},
	\end{equation}
	the continuity of the free energy across the $CB$ phase boundary imposes the second constraint\cite{yipThermodynamicConsiderationsPhase1991}:
	\begin{equation}
		\frac{1 - p_2/p_4}{1 - p_2/p_1} = \frac{\mp (r_1/r_4)r_3 - r_1^2}{1 - r_1^2}.
	\end{equation}
	From the phase boundaries measured in this work, we have the slopes $p_1 = 2.9$ K/GPa, $p_2 = p_3 = -1.08$ K/GPa, and $p_4 = -77.9$ K/GPa. Using these, we get for the left hand side $\frac{1 - p_2/p_4}{1 - p_2/p_1} = 0.71$. We use the quantitative specific heat values of Vasina \textit{et al.}  \cite{vasinaQuantitativeThermodynamicStudy2026} to obtain $r_3 = 0.984$ and $r_1 = 0.027$, which gives us
	\begin{equation}
		\frac{r_1^2}{r_4^2} = \frac{\alpha_{AN}}{\alpha_{CB}} = 0.52
		\label{eq:ratios}
	\end{equation}
	That is, the specific heat jump across the $CB$ (\Tcstar) phase boundary must be approximately twice the specific heat jump across the $AN$ (\Tct) phase boundary. Although we do not measure the specific heat jump at \Tcstar, the Ginzburg Landau theory we construct gives us $\alpha_{CB}\ll\alpha_{BN}$ and $\alpha_{AN}\ll\alpha_{BN}$, consistent with this constraint.
	
	Using $r_4=(r_1^2 + r_3^2 -1)^{1/2}$ in \autoref{eq:yip1} gives us the final constraint:
	\begin{equation}
		\frac{1 - p_1/p_3}{1-p_1/p_2} = \frac{r_3 \pm r_1r_4}{(1-r_1^2)r_3}.
	\end{equation}
	Since we measure $p_2 = p_3$ in this work, the left hand side of this constraint $\frac{1 - p_1/p_3}{1-p_1/p_2}=1$. From the values of $r_1$, $r_3$ and \autoref{eq:ratios}, this gives us for the right hand side $\frac{r_3 \pm r_1r_4}{(1-r_1^2)r_3} = 0.9997$, satisfying the final constraint.
	
	\subsection{Symmetry Analysis}
	
	The application of a $b$-axis magnetic field reduces the point group symmetry of \ute from D$_{2\text{h}}$ to C$_{2\text{h}}$, removing the two mirror planes containing the magnetic field vector. 
	
	The fact that we still see second-order phase transitions between the SC1 and SC2 superconducting states in \autoref{fig:intro} allows us to constrain the possible symmetries of the SC1+SC2 order in \ute. In particular, SC1 and SC2 must belong to separate irreducible representations of the reduced C$_{2\text{h}}$ symmetry to allow for a second-order phase transition between them. In \autoref{tab:d2h_c2h}, we show the mapping of each D$_{2\text{h}}$ irrep. when a symmetry breaking $b$ axis field is applied. We can then rule out the $A_u + B_{2u}$ and $B_{1u} + B_{3u}$ combinations (written in the $D_{2h}$ nomenclature) among the odd-parity irreps., since both components map onto the same irrep. in C$_{2\text{h}}$ and therefore any transition between these two states would generically be first order.
	
	Additionally, the lack of a jump in \cff at $P>\Ps$ also constrains certain combinations. Since \cff measures the coupling of the $\epsilon_{xz}$ strain to the order parameter, order parameters combinations that couple linearly to  $\epsilon_{xz}$ may also be ruled out \cite{theussSinglecomponentSuperconductivityUTe22024,ghoshThermodynamicEvidenceTwocomponent2021}. This also rules out the $A_u + B_{2u}$ and $B_{1u} + B_{3u}$ combinations among odd-parity representations. 
	
	
	\subsection{Zero Pressure Phase Diagram}
	
	Although we do not apply magnetic fields large enough to reach the multi-component state at ambient pressure, the $B_b-T-P$ phase diagram of Figure 4 of the main text allows us to constrain the $P=0$ phase diagram as a function of temperature and $b$-axis magnetic field, shown in Extended Data Figure 4. The multi-component state is accessible well within the reach of many commercially available superconducting magnets at fields over 12 T --- enabling studies with more conventional techniques without the need of a pressure cell.
	
	\subsection{Data Used for Zero-Field Phase Diagram Construction}
	
	The full set of data we measure to construct the Pressure-Temperature phase diagram at zero field in Figure 2 of the main text are shown in Extended Data Figure 5. Superconducting transitions manifest as jumps in the compressional elastic constants \coo and \ctt, and as kinks in the shear elastic constant \cff. 
	
	The \ctt data shown here (Extended Data Figure 5b) follow the same trend as \coo shown in the main text. We see a single transition as a downward jump at \Tco when $P<\Ps$. For $P>\Ps$ the \Tct transition appears as a large downward jump, while the \Tco transition is too small to be seen on the scale of Extended Data Figure 5b (Extended Data Figure 6a and b show \ctt data at 0.36 and 0.77 GPa on a larger scale, with insets clearly showing the downward jumps at \Tco). At $P = 0.21$ GPa, we see an upward jump at \Tcstar. Note that there is a subtle kink about 50 mK below \Tcstar, which we attribute to pressure inhomogeniety within the sample.  We attribute this to either small internal inhomogeneities or small external pressure inhomogeneities in the sample: because \Tcstar drops by nearly 1.5 K over a 0.01 GPa pressure range, even part-per-thousand inhomogenieties in the pressure can lead to smearing of \Tcstar over tens of mK.
	
	The \cff data are shown in (Extended Data Figure 5c), and exhibit kinks at the \Tco and \Tct superconducting transitions.
	
	\subsection{Absence of Hysteresis at \Tcstar}
	
	To demonstrate that the \Tcstar transition measured in this work is second-order, we measure \ctt and \att while sweeping temperature up and down through \Tcstar. These data are shown in Extended Data Figure 6c and d. Note the absence of any hysteresis.
	
	\subsection{Data Used for Field-Temperature Phase Diagram Construction}
	
	The full dataset used to construct the field--temperature phase diagrams in \autoref{fig:infield} is shown in Extended Data Figure 7 and Extended Data Figure 8. Temperature sweeps of \ctt\ and \cff\ near the tetracritical point are presented in Extended Data Figure 7, while Extended Data Figure 8 shows the corresponding field sweeps of \ctt\ and \att\ at the same pressures.
	
	\subsection{Constraining the \Tcstar Phase Boundary in \BPT space}
	
	We obtain the \Tcstar phase boundary shown in \autoref{fig:zerofield} by constraining a sheet of \Tcstar phase transitions in \BPT space and then projecting it onto $B_b = 0$. This is a stronger constraint than using the transitions we observe at $B_b = 0$ alone. We show this phase boundary along with all the \Tcstar transitions observed in this work in Extended Data Figure 9.
	
	\section{Data Availability}
	
	Data measured in this work are freely available at \url{https://github.com/CHiLL-Ramshaw/manuscripts-supporting_data/tree/main/2026_UTe2_Tetracriticality}.
		
		\begin{figure}
				\centering
				\includegraphics[width=1\linewidth]{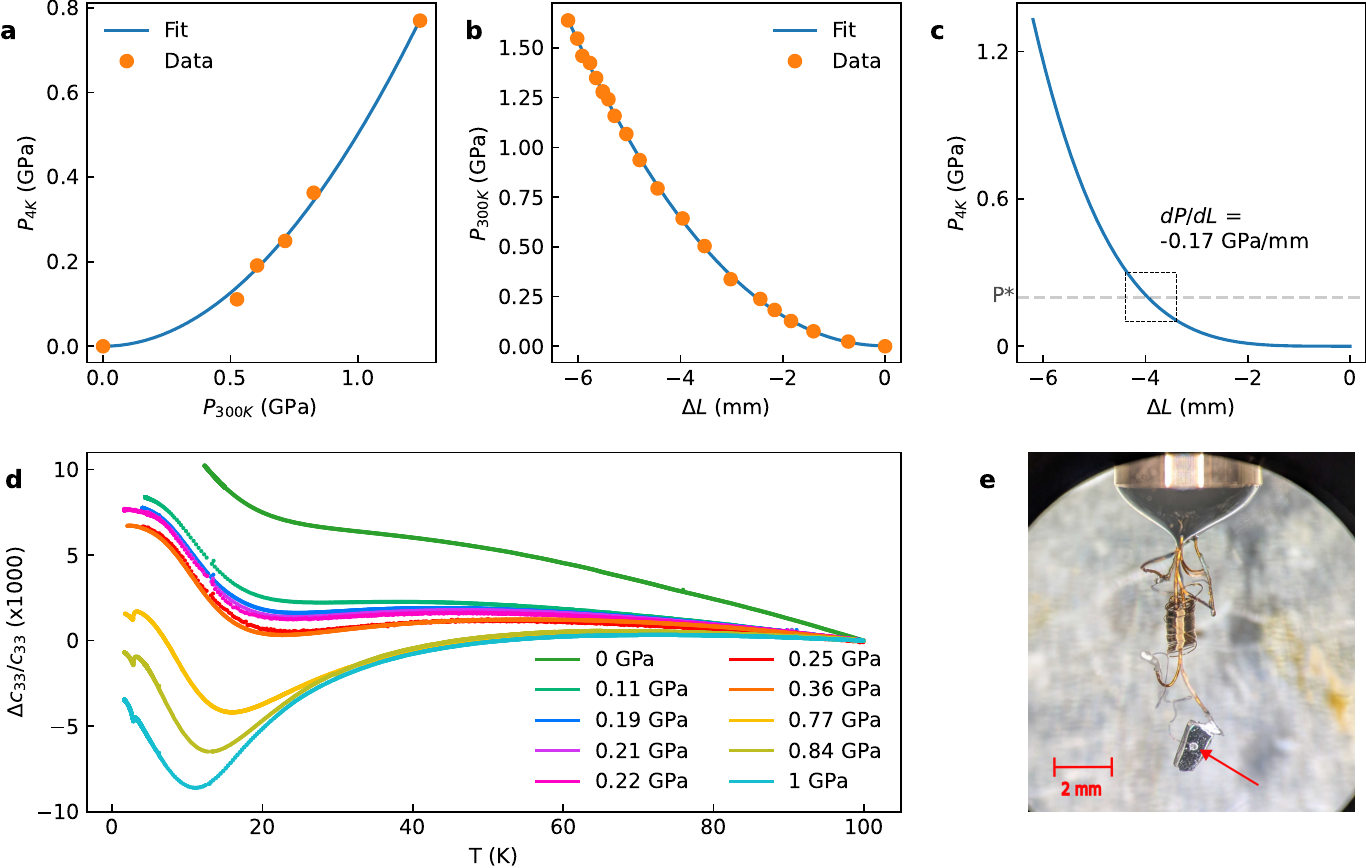}
				\caption{\textbf{Accurately tuning the pressure at low temperature.} a) The pressure of the sample at low temperature (4 K) as measured by \Tco and \Tct versus the pressure as measured by a manganin manometer at room temperature. We fit the data to a second order polynomial. b) The pressure measured at room temperature versus the change in length of the pressure cell, proportional to the change in volume of the pressure medium. We fit the data to a fourth order polynomial. c) Combining the fits in (a) and (b), the predicted low-temperature pressure versus the pressure cell length. The slope of this curve in the region close to the tetracritical point (dashed box) is used to accurately change the pressure of the sample by changing the length of the pressure cell. d) \ctt data up to 100 K, used as a consistency check as we tune pressure. The minimum in the data occurs due to the Kondo effect and is highly sensitive to pressure. e) Sample when mounted inside the pressure cell. Seen are the coil of manganin wire used as a manometer at room temperature, along with the sample of \ute. The pulse-echo transducer is indicated by a red arrow and is seen as a small silver dot on the sample. Because of its small size ($\sim100\xspace\mu$m), we measure a small sample volume --- given by the area of the transducer ($\sim \pi (100\xspace\mu \text{m})^2$) times the thickness of the sample ($\sim 300 \mu\xspace$m). This is much smaller than in other techniques, leading to a much lower pressure inhomogeniety.}
				\label{fig:pressurization}
			\end{figure}
		
		\begin{figure}
				\centering{}
				\includegraphics[width=\linewidth]{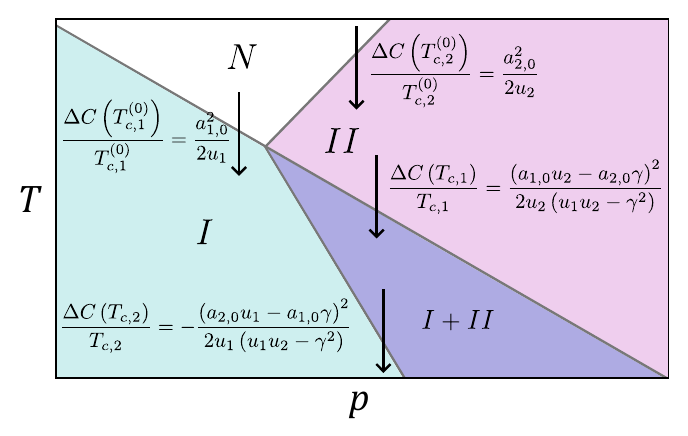}
				\caption{\textbf{Phase Diagram and Heat Capacity Jumps.} The temperature and pressure axes are denoted by $T$ and $p$ respectively. The SC1, SC2 and SC1+SC2 states are denoted by I,II and I+II respectively. Jumps in $\Delta C/T$ occur at each phase boundary, with arrows indicating the crossing direction.}
				\label{fig:theory_diag}
			\end{figure}
		
		\begin{figure}
				\centering
				\includegraphics[width=0.6\linewidth]{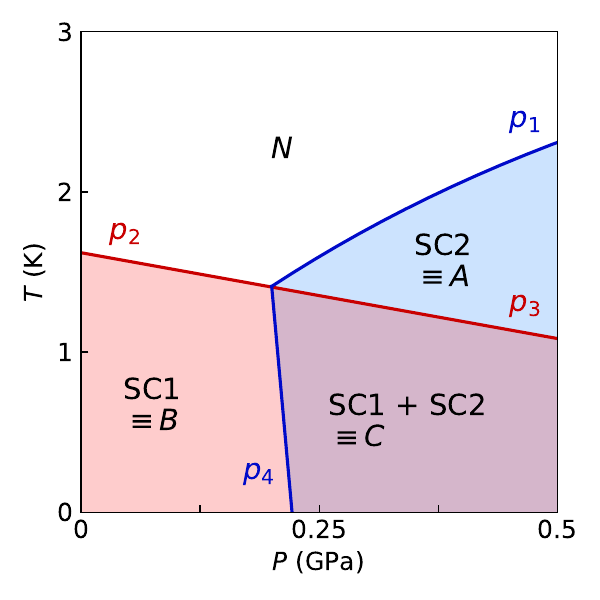}
				\caption{\textbf{The $P-T$ phase diagram with definitions used in \hyperref[sec:yip]{General Thermodynamic Constraints}}. The normal state, SC2, SC2 and SC1+SC2 phases are referred to as N, A, B and C respectively, following the notation used in ref. \cite{yipThermodynamicConsiderationsPhase1991}. The slopes of the $AN$,$BN$,$CA$ and $CB$ phase boundaries are defined to be $p_1$, $p_2$, $p_3$ and $p_4$ respectively.}
				\label{fig:yip}
			\end{figure}

		\begin{table}
				\centering
				\begin{tabular}{cc}
						\hline
						$D_{2h}$ irrep. & $C_{2h}$ irrep. \\
						\hline
						$A_g$    & $A_g$ \\
						$B_{1g}$ & $B_g$ \\
						$B_{2g}$ & $A_g$ \\
						$B_{3g}$ & $B_g$ \\
						$A_u$    & $A_u$ \\
						$B_{1u}$ & $B_u$ \\
						$B_{2u}$ & $A_u$ \\
						$B_{3u}$ & $B_u$ \\
						\hline
					\end{tabular}
				\caption{A mapping between $D_{2h}$ and $C_{2h}$ irreps. in \ute.}
				\label{tab:d2h_c2h}
			\end{table}

		\begin{figure}
				\centering
				\includegraphics[width=0.6\linewidth]{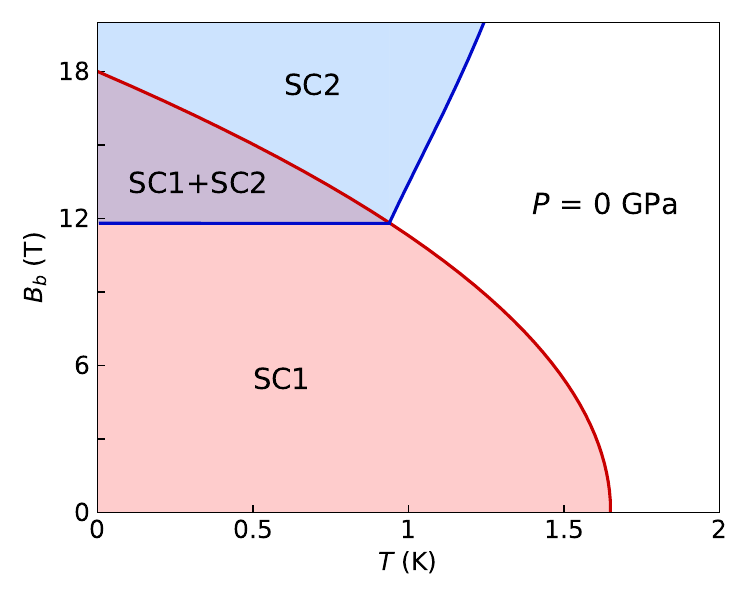}
				\caption{\textbf{Ambient pressure phase diagram.} The $B_b-P-T$ phase diagram we construct in Figure 4 of the main text constrains the zero-pressure phase boundaries as a function of magnetic field and temperature. SC1 and SC2 phases are shaded red and blue respectively. The SC1+SC2 multi-component state, shaded purple, is accessible between 12 and 18 T at ambient pressure. The \Tco and \Tct phase boundaries we observe agree quantitatively with previous measurements of the phase diagram \cite{sakaiFieldInducedMultiple2023,rosuelFieldInducedTuningPairing2023}.}
				\label{fig:extrapolated}
			\end{figure}

		\begin{figure}
				\centering
				\includegraphics[width=\linewidth]{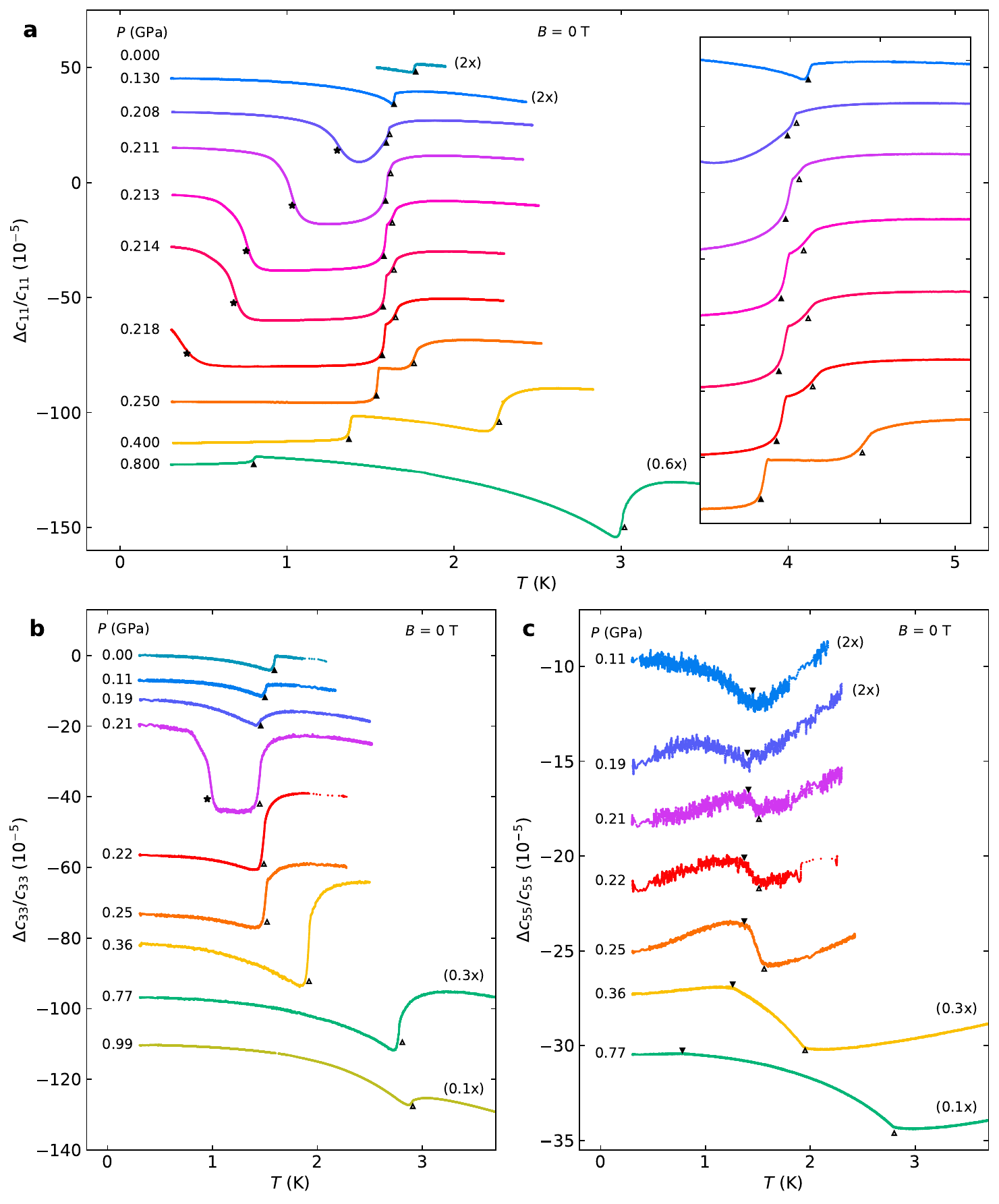}
				\caption{\textbf{Full set of \coo,\ \ctt and \cff data used to construct the zero-field $P-T$ phase diagram.} (a) \coo data. The inset shows zoomed in \coo at pressures near \Ps. (b) \ctt data (c) \cff data, measured simultaneously with the \ctt data. \Tco, \Tct and \Tcstar are marked by solid triangles, hollow triangles and stars respectively. }
				\label{fig:all_zero_field}
			\end{figure}
		
		\begin{figure}
				\centering
				\includegraphics[width=0.8\linewidth]{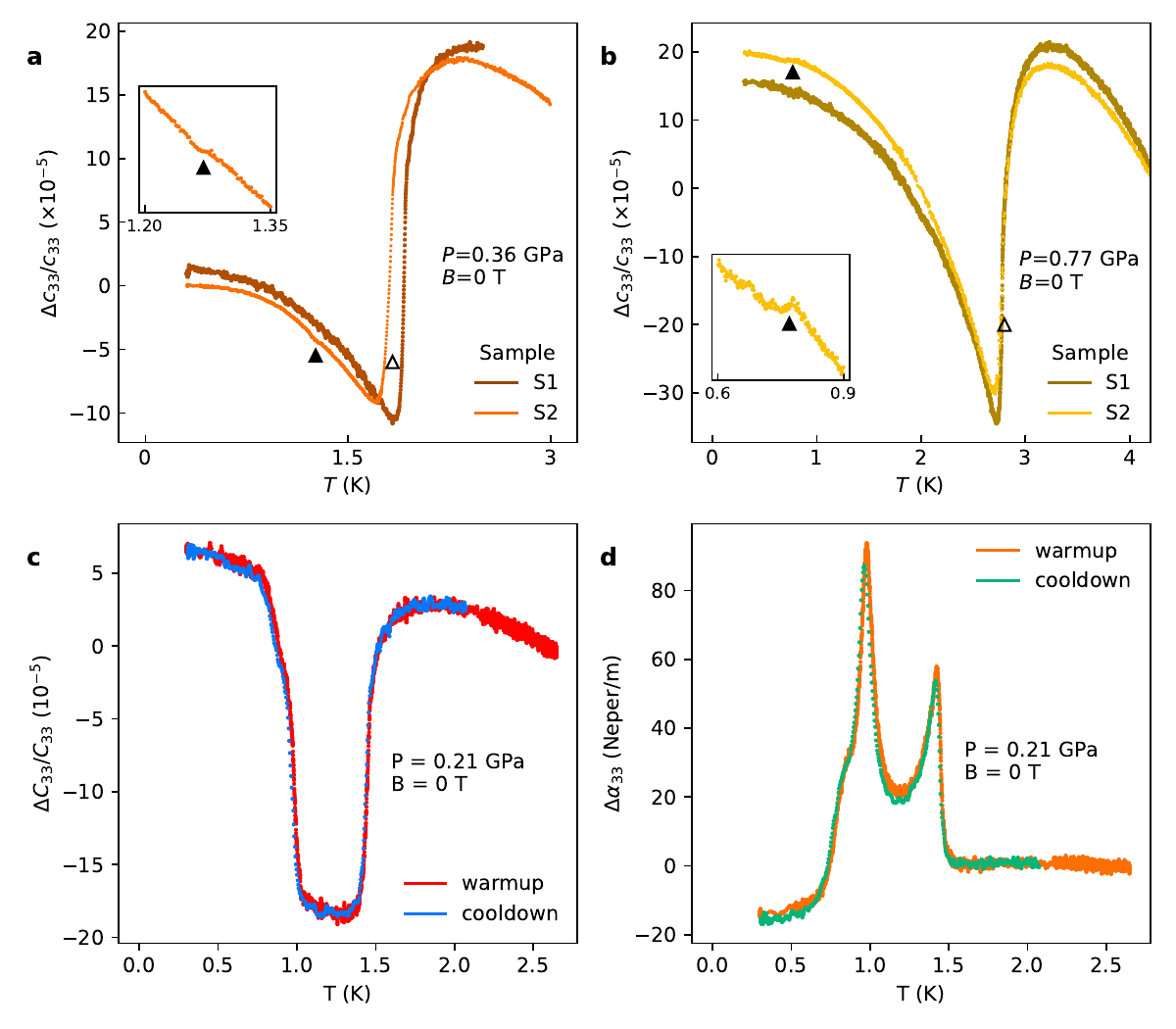}
				\caption{\textbf{a) and b) Jumps in \ctt at \Tco}. Although not visible on the scale of Extended Data Figure 5, we observe small downward jumps as the sample enters the SC1 state at \Tco for pressures  $P>\Ps$, marked by filled triangles. Note that these jumps are measured on a different sample S1 optimized for 
						\ctt measurement --- \ctt data at 0.36 and 0.77 GPa in Extended Data Figure 5 is sample S2. Insets: Data zoomed in on the transition at \Tco. \textbf{c) and d) Absence of hysteresis at \Tcstar}. \ctt (c) and \att (d) as a function of temperature, measured on cooling and warming the sample through \Tcstar. The absence of any hysteresis confirms that \Tcstar is a second order phase transition.}
				\label{fig:twoTcs_plus_hysteresis}
			\end{figure}

		\begin{figure}
				\centering
				\includegraphics[width=\linewidth]{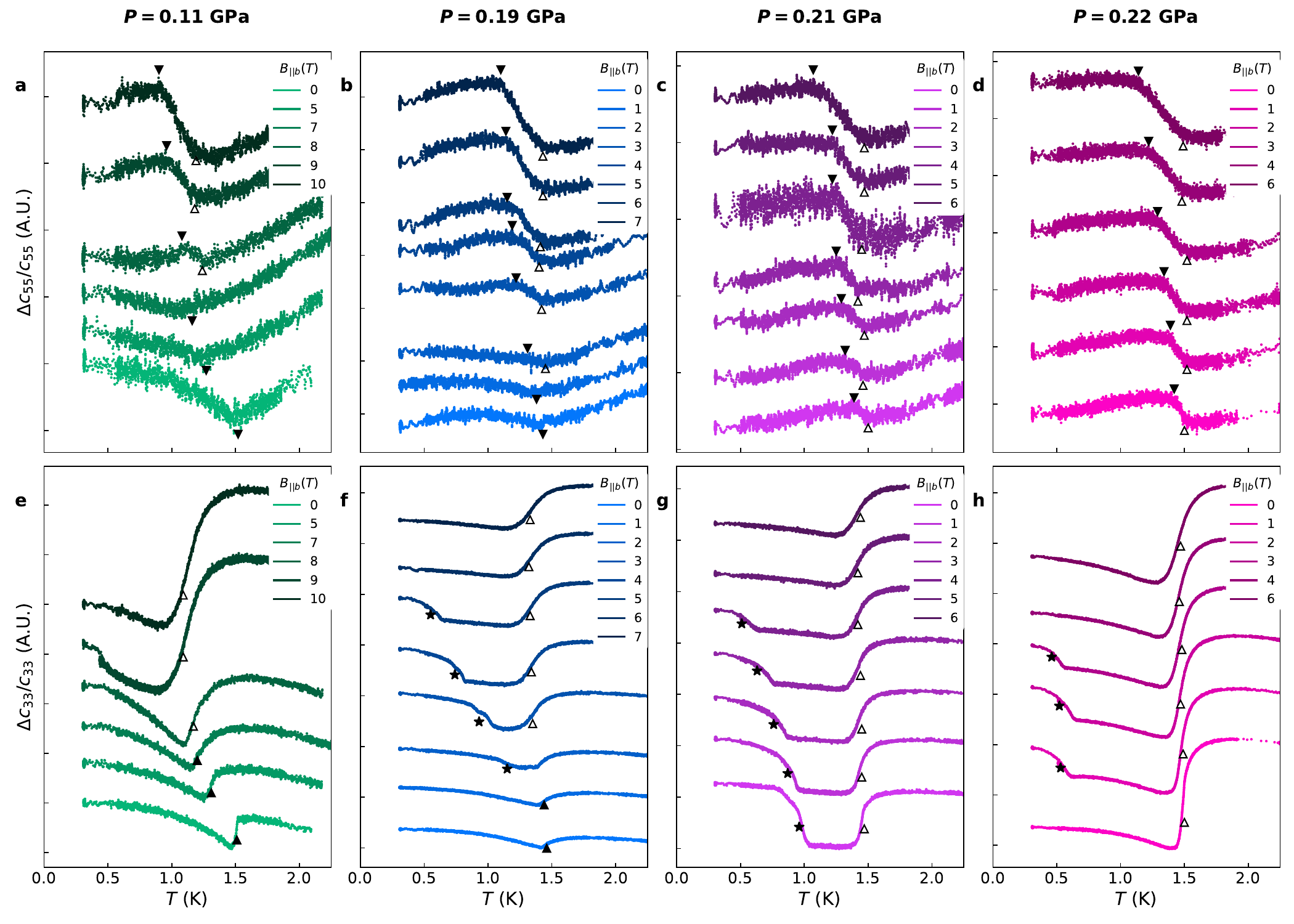}
				\caption{\textbf{Full set of \ctt and \cff data measured as a function of temperature.} (a-d) \cff at different pressures and magnetic fields, with kinks corresponding to the \Tco and \Tct transitions marked by solid and hollow triangles respectively. e-h) \ctt at different pressures and magnetic fields. \Tco, \Tct and \Tcstar are marked by solid triangles, hollow triangles and stars respectively. }
				\label{fig:extended_tempsweeps}
			\end{figure}
		
		\begin{figure}
				\centering
				\includegraphics[width=\linewidth]{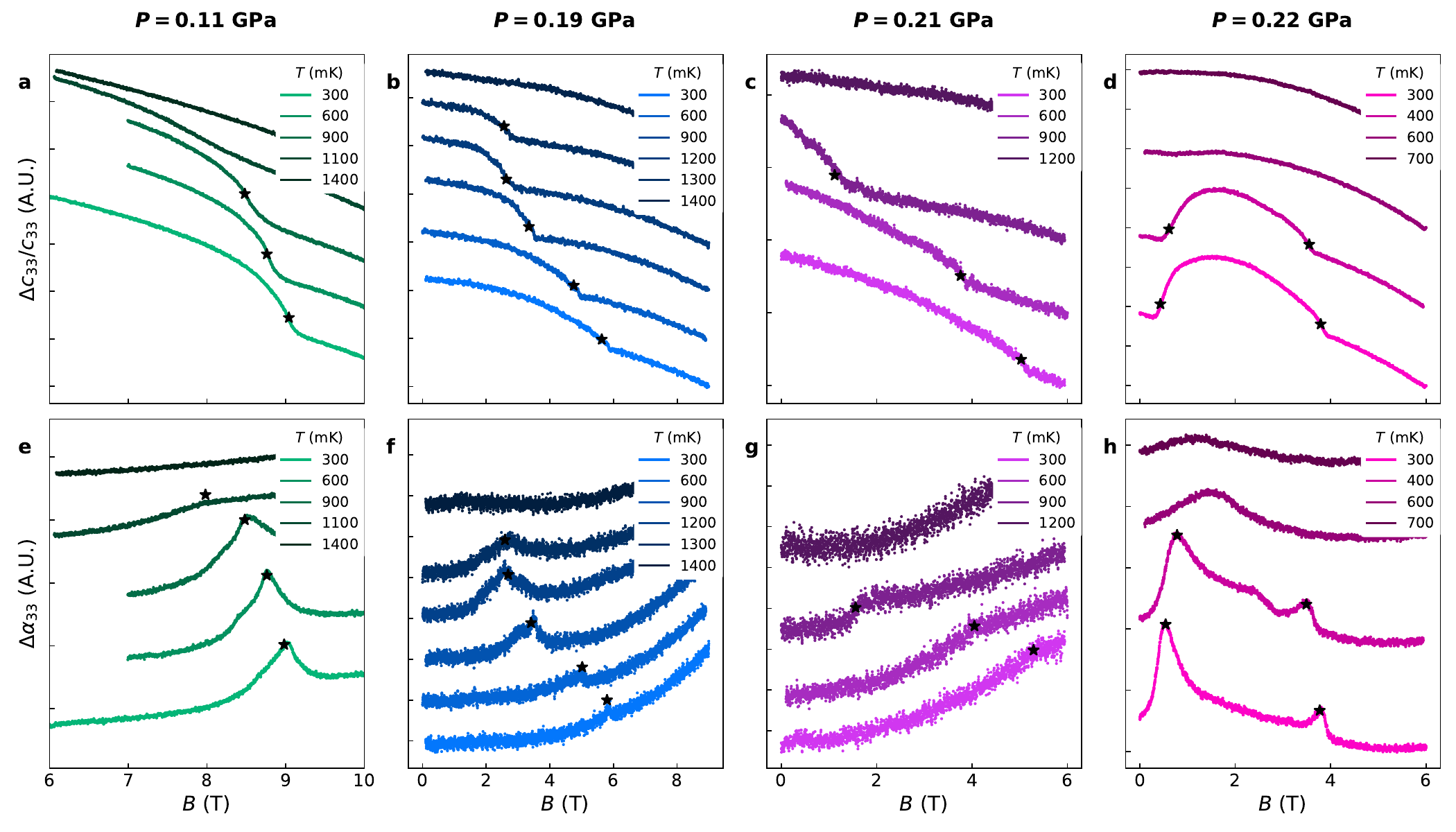}
				\caption{\textbf{Full set of \ctt and \att data measured as a function of $b$-axis magnetic field.} a-d) \ctt at different temperatures and pressures. We observe a broad downward jump as the sample enters the SC2 state, marked by a solid black star. e-h) \att data at different temperatures and pressures. We see an order parameter relaxation peak at the SC2 phase boundaries, marked by a solid black star.}
				\label{fig:extended_fieldsweeps}
			\end{figure}

		\begin{figure}
				\centering
				\includegraphics[width=0.8\linewidth]{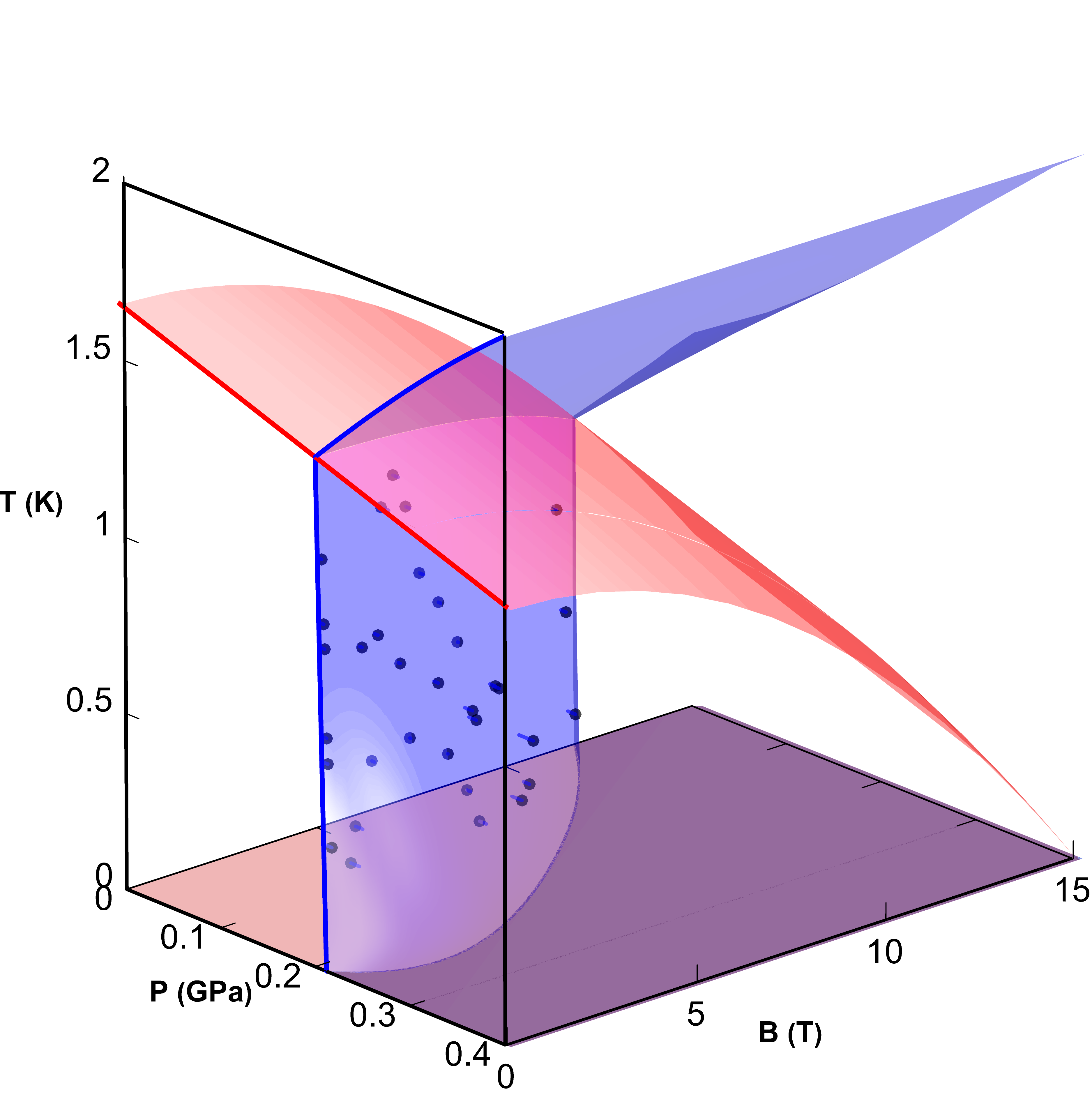}
				\caption{\textbf{The \Tcstar phase boundary constructed in this work.} All the \Tcstar transitions we measure in this work are shown as solid black points, connected to the \Tcstar phase boundary (shown as a blue sheet) by thin blue lines. We obtain the zero field phase boundary via the intersection of this sheet with the $B_b = 0$ plane (shown as a black frame). }
				\label{fig:sheet}
			\end{figure}

	\renewcommand{\refname}{Methods References}

\end{document}